\documentclass[prl,onecolumn,notitlepage,11pt,tightenlines,amsmath,amssymb,showpacs,superscriptaddress,nofootinbib,a4paper,longbibliography]{revtex4-1}

\allowdisplaybreaks 

\usepackage[T1]{fontenc}
\usepackage{lmodern}
\usepackage{fixltx2e}
\usepackage{microtype} 
\usepackage{xspace}
\usepackage{enumitem}

\makeatletter
\newcommand\etc{etc\@ifnextchar.{}{.\@}\xspace}

\makeatother

\usepackage{graphicx}
\graphicspath{{image/}}

\newcommand{\inlinegraphic}[2]{
  \dimendef\grafheight=255\dimendef\grafvshift=254
  \grafheight=#1
  \grafvshift=-0.5\grafheight
  \advance\grafvshift by 0.5ex
  \raisebox{\grafvshift}{\includegraphics[height=\grafheight]{images/#2}\xspace}
}

\newcommand{\ninlinegraphic}[2][1.0]{
  \dimendef\grafheight=255\dimendef\grafvshift=254
  \setbox0 = \hbox{\scalebox{#1}{\includegraphics{images/#2}}}
  \grafheight=\the\ht0
  \grafvshift=-0.5\grafheight
  \advance\grafvshift by 0.5ex
  \raisebox{\grafvshift}{\includegraphics[height=\grafheight]{images/#2}\xspace}
}

\usepackage{xcolor}
\usepackage{color}
\definecolor{Blue}{rgb}{0.00, 0.00, 1.00}
\definecolor{Red}{rgb}{1.00, 0.00, 0.00}
\definecolor{Green}{rgb}{0.00, 1.00, 0.00}

\definecolor{codegreen}{rgb}{0,0.6,0}
\definecolor{codegray}{rgb}{0.5,0.5,0.5}
\definecolor{codepurple}{rgb}{0.58,0,0.82}
\definecolor{backcolour}{rgb}{0.95,0.95,0.92}

\usepackage{latexsym}
\usepackage{amssymb}
\usepackage{amsmath} 
\usepackage{amsfonts}
\usepackage{amsthm}
\usepackage{stmaryrd}
\usepackage{mathtools}

\usepackage{amsthm}

\theoremstyle{definition}
\theoremstyle{definition}
\theoremstyle{definition}
\theoremstyle{definition}
\theoremstyle{definition}
\theoremstyle{definition}
\theoremstyle{definition}

\newcommand{\sizeof}[1]{
  \left|#1\right|}

\DeclareMathOperator*{\argmin}{arg\,min}

\usepackage{diagrams} 
\newarrow{Equals}{}{=}{}{=}{}

\ifx\numericids\undefined

\else

\fi

\usepackage{dsfont}

\newcommand{\tket}{\ensuremath{\mathsf{t}|\mathsf{ket}\rangle}\xspace}

\usepackage{listings}    
\lstdefinestyle{mystyle}{
  backgroundcolor=\color{backcolour},   
  commentstyle=\color{codegreen},
  keywordstyle=\color{magenta},
  numberstyle=\tiny\color{codegray},
  stringstyle=\color{codepurple},
  basicstyle=\footnotesize,
  breakatwhitespace=false,         
  breaklines=true,                 
  captionpos=b,                    
  keepspaces=true,                 
  numbers=left,                    
  numbersep=5pt,                  
  showspaces=false,                
  showstringspaces=false,
  showtabs=false,                  
  tabsize=2
}
     
\usepackage{cases}
\usepackage{bm}
\usepackage{braket}
\usepackage{changepage}
\usepackage{longtable}
\usepackage[caption=false]{subfig}
\usepackage{wrapfig}
\usepackage{commath}
\usepackage{dsfont}

\usepackage{makecell}
\usepackage{csvsimple}
\usepackage{xstring}

\usepackage{siunitx}
\newcommand{\optnum}[2]{\IfInteger{#1}{\ifnum0=0#1\relax0*\else\num{#1}\fi}{\IfDecimal{#1}{\num{#1}}{#2}}}

\usepackage{hyperref}
\hypersetup{
  colorlinks=true,      
  linkcolor=blue,       
  citecolor=magenta,    
  filecolor=red,        
  urlcolor=cyan         
}

\usepackage{tikzcircuits}

\usetikzlibrary{graphs,graphs.standard,shapes.misc}
\tikzset{cross/.style={cross out, draw=black, minimum size=2*(#1-\pgflinewidth), inner sep=0pt, outer sep=0pt}, cross/.default={3.5pt}}

\usepackage{enumitem}
\setlist{noitemsep}

\begin{document}
    
\title{On the qubit routing problem}
\author{Alexander Cowtan}
\affiliation{Cambridge Quantum Computing Ltd, 9a Bridge Street, Cambridge, CB2 1UB, UK}
\author{Silas Dilkes}
\affiliation{Cambridge Quantum Computing Ltd, 9a Bridge Street, Cambridge, CB2 1UB, UK}
\author{Ross Duncan}
\email{ross.duncan@cambridgequantum.com}
\affiliation{Cambridge Quantum Computing Ltd, 9a Bridge Street, Cambridge, CB2 1UB, UK}
\affiliation{University of Strathclyde, 26 Richmond Street, Glasgow,  G1 1XH, UK}
\author{Alexandre Krajenbrink}
\email{alexandre.krajenbrink@cambridgequantum.com}
\affiliation{Cambridge Quantum Computing Ltd, 9a Bridge Street, Cambridge, CB2 1UB, UK}
\affiliation{Laboratoire de Physique de l'\'Ecole Normale   Sup\'erieure, PSL University, CNRS, Sorbonne Universit\'es, 24 rue   Lhomond, 75231 Paris Cedex 05, France}
\author{Will Simmons}
\affiliation{Cambridge Quantum Computing Ltd, 9a Bridge Street, Cambridge, CB2 1UB, UK}
\author{Seyon Sivarajah}
\affiliation{Cambridge Quantum Computing Ltd, 9a Bridge Street, Cambridge, CB2 1UB, UK}
    
\date{\today}

\begin{abstract}
  We introduce a new architecture-agnostic methodology for mapping
  abstract quantum circuits to realistic quantum computing devices
  with restricted qubit connectivity, as implemented by Cambridge
  Quantum Computing's \tket compiler.  We present empirical results
  showing the effectiveness of this method in terms of reducing 
  two-qubit gate depth and two-qubit gate count, compared to other
  implementations.
\end{abstract}

\maketitle

\section{Introduction}
    
\begin{figure}[h!]
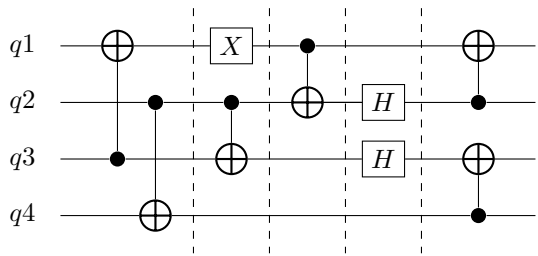

  \ctikzfig{rd-circuit-example}
  \caption{Example of a quantum circuit containing one and two-qubit gates acting on four qubits, $q1$, $q2$, $q3$ and $q4$. This circuit has five time steps, each with gates acting on disjoint sets of qubits. \label{fig:circuit}}
\end{figure}

There is a significant gap between the theoretical literature on
quantum algorithms and the way that quantum computers are implemented.
The simple and popular \emph{quantum circuit model} presents the
quantum computer as a finite number of qubits upon which quantum gates
act; see Fig.~\ref{fig:circuit} for an example. Typically gates act on
one or two qubits at a time, and the circuit model allows multi-qubit
gates to act on any qubits without restriction. However, in realistic
hardware the qubits are typically laid out in a fixed two or three
dimensional topology where gates may only be applied between
neighbouring qubits.  In order for a circuit to be executed on such
hardware, it must be modified to ensure that whenever two qubits are
required to interact, they are adjacent in memory.  This is a serious
departure from the von Neumann architecture of classical computers,
where operations may involve data at distant locations in memory
without penalty.

We refer to the task of modifying a circuit to conform to the memory
layout of a specific quantum computer as the \emph{qubit routing
  problem}.  When non-adjacent qubits are required to interact we
can insert additional SWAP gates to exchange a qubit with a neighbour,
moving it closer to its desired partner.  In general many -- or even
all -- of the qubits may need to be swapped, making this problem
non-trivial.  Since quantum algorithms are usually designed without
reference to the connectivity constraints of any particular hardware,
a solution to the routing problem is required before a quantum circuit
can be executed.  Therefore qubit routing forms a necessary stage of
any compiler for quantum software.  Current quantum computers -- the
so-called NISQ\footnote{``Noisy intermediate-scale quantum'' devices;
  see \cite{preskill2018quantum} for a survey.} devices -- impose
additional constraints.  Their short coherence times and relatively
low fidelity gates require that the circuit depth and the total number
of gates are both as low as possible.  As routing generally introduces
extra gates into a circuit, increasing its size and depth, it is
crucial that the circuit does not grow too much, or its performance
will be compromised.

The general case of the routing problem, also called the qubit
allocation problem, is known to be infeasible.  The sub-problem of
assigning logical qubits to physical ones is equivalent to sub-graph
isomorphism \cite{siraichi2018qubit}, while determining the optimal
swaps between assignments is equivalent to token-swapping
\cite{10.1007/978-3-319-07890-8_31} which is at least \textsc{np}-hard
\cite{Bonnet2018} and possibly \textsc{pspace}-complete
\cite{JERRUM1985265}.  Siraichi et al. \cite{siraichi2018qubit} propose
an exact dynamic programming method (with complexity exponential in
the number of qubits) and a heuristic method which approximates it
well, at least on the small (5 qubit) circuits considered.  Zulehner
et al. \cite{Zulehner:2017aa} propose an algorithm based on depth
partitioning and A* search which is specialised to the architectures
of IBM devices \cite{ibm_doc_tokyo}.  Other approaches take advantage
of the restricted topology typically found in quantum memories such as
linear nearest neighbour \cite{hirata:2011:linear} or hypercubic
\cite{brierley2015efficient} which rely on classical sorting networks;
see Appendix \ref{sec:classical_sorting} for a discussion of this
approach.  Lower bound results for routing are presented by 
Herbert~\cite{herbert2018depth}.


In this paper we describe the solution to the routing problem
implemented in \tket, a platform-independent compiler developed by
Cambridge Quantum Computing Ltd\footnote{\tket is available as a
  python module from \url{https://pypi.org/project/pytket/}.}. The
heuristic method in \tket matches or beats the results of other
circuit mapping systems in terms of depth and total gate count of the
compiled circuit, and has much reduced run time allowing larger
circuits to be routed.

Aside from qubit routing, \tket also provides translation from general
circuits to any particular hardware-supported gate set, a variety of
advanced circuit optimisation routines, and support for most of the
major quantum software frameworks.  These will be described in future
papers.  Compilation through \tket guarantees hardware compatibility
and minimises the depth and gate count of the final circuit across a
range of hardware and software platforms.

In Section~\ref{sec:routing-problem} we formalise the problem and
present an example instance.  In Section~\ref{sec:algo} we describe
the method used by \tket to solve the problem.   In
Section~\ref{sec:graph_repre_qc} we describe some of the architectures
on which we tested the algorithm and in Section~\ref{sec:results} we
present empirical results of \tket's performance, both in terms of
scaling and in comparison to other compiler software.  Full tables of
results are provided in the Appendix.







\section{The routing problem}
\label{sec:routing-problem}

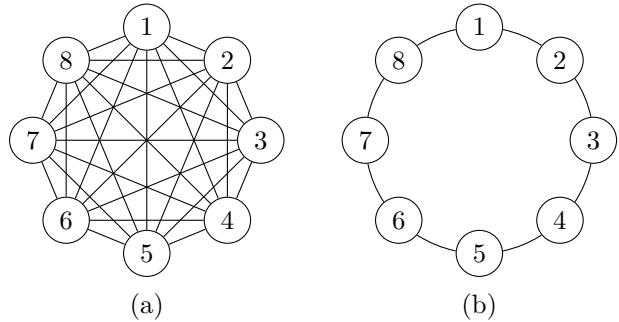
\begin{figure}[t!]
  \begin{center}
    \begin{tikzpicture}
  \def \radiuss {2.2cm} \graph [nodes={draw,circle}]{ subgraph K_n
    [n=8,clockwise,radius=1.5cm] }; \node at ({-90}: \radiuss)
  {(a)};
\end{tikzpicture}
    \hspace{0.4cm}
    \begin{tikzpicture}
  \def \n {8} \def \radius {1.5cm} \def \radiuss {2.2cm} \def
  \margin {12} 
  
  \foreach \s in {1,...,\n} { \node [draw, circle] at ({-360/\n *
      (\s - 1)+90}:\radius) {$\s$}; \draw[ -, >=latex] ({360/\n *
      (\s - 1)+\margin}:\radius) arc ({360/\n * (\s -
      1)+\margin}:{360/\n * (\s)-\margin}:\radius); } \node at
  ({-90}: \radiuss) {(b)};
\end{tikzpicture}
  \end{center}
  \vspace*{-1em}
  \caption{Nodes in the graph represent physical qubits
    and edges are the allowed interactions. (a) The circuit model
    assumes all-to-all communication between qubits, \textit{i.e.} a complete
    graph and (b) a physically realistic one-dimensional nearest
    neighbour cyclic graph, the
    ring.\label{fig:complete}}
\end{figure}

We represent a quantum computer as a graph where nodes are physical
qubits and edges are the allowed 2-qubit interactions\footnote{We don't
consider architectures with multi-qubit interactions involving more
than two qubits.}.  Since the circuit model assumes we can realise a
two-qubit gate between any pair of qubits, it is equivalent to the
complete graph (Fig.~\ref{fig:complete}a).  Realistic qubit
architectures are connectivity limited: for instance, in most
superconducting platforms the qubit interaction graph must be planar;
ion traps present more flexibility, but are still not fully connected.
For now we will use the ring graph (Fig. \ref{fig:complete}b) as a
simple example.  Given such a restricted graph, our goal is to emulate
the complete graph with minimal additional cost.
 
From this point of view, the routing problem can be stated as follows.
Given \textit{(i)} an arbitrary quantum circuit and \textit{(ii)} a
connected graph specifying the allowed qubit interactions, we must
produce a new quantum circuit which is equivalent to the input
circuit, but uses only those interactions permitted by the
specification graph.  Provided the device has at least as many qubits
as the input circuit then a solution always exists; our objective is
to minimise the size of the output circuit.

\subsection{Example: routing on a ring}
\label{sec:routing-example}

Let's consider the problem of routing the circuit shown in
Fig.~\ref{fig:circuit} on the ring graph of
Fig.~\ref{fig:complete}(b).  The first step is to divide the circuit
into \emph{timesteps}, also called \emph{slices}.  Loosely speaking, a
timestep consists of a subcircuit where the gates act on disjoint sets
of qubits and could in principle all be performed simultaneously (see
Section~\ref{subsec:packing} for a precise definition).  The single
qubit gates have no bearing on the routing problem so can be ignored,
and thus a timestep can be reduced to a set of qubit pairs that are
required to interact via some 2-qubit gate.

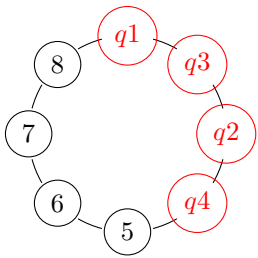
\begin{figure}[h!]
  \begin{tikzpicture}
    
  \def \n {8} 
  \def \radius {1.3cm} 
  \def \radiuss {2cm} 
  \def \margin {15} 

  \foreach \s in {1,...,\n} 
  {
    \draw[ -, >=latex] ({360/\n * (\s -         1)+\margin}:\radius) arc ({360/\n * (\s -       1)+\margin}:{360/\n * (\s)-\margin}:\radius); }

  \node [draw, circle, red] at ({-360/\n * (1 - 1)+90}:\radius) {$q1$};
  \node [draw, circle, red] at ({-360/\n * (2 - 1)+90}:\radius) {$q3$};
  \node [draw, circle, red] at ({-360/\n * (3 - 1)+90}:\radius) {$q2$};
  \node [draw, circle, red] at ({-360/\n * (4 - 1)+90}:\radius) {$q4$};
  \node [draw, circle] at ({-360/\n * (5 - 1)+90}:\radius) {$5$};
  \node [draw, circle] at ({-360/\n * (6 - 1)+90}:\radius){$6$};
  \node [draw, circle] at ({-360/\n * (7 - 1)+90}:\radius) {$7$};
  \node [draw, circle] at ({-360/\n * (8 -1)+90}:\radius) {$8$}; 
\end{tikzpicture}
  \caption{An initial mapping of logical qubits to nodes. Highlighted
    nodes are labelled with the mapped qubit. \label{fig:initial_mapping_3}}
\end{figure}

Next, the logical qubits of the circuit must be mapped to the nodes of
the graph.  For our example a reasonable initial mapping is
$q1\rightarrow1$, $q3\rightarrow2$, $q2\rightarrow3$, $q4\rightarrow4$
as shown in Fig.~\ref{fig:initial_mapping_3}.  This has the advantage
that the qubits which interact in the first timestep are adjacent in
the graph, and the same for the second timestep.

\begin{figure}[t!]
  \begin{center}
          \begin{tikzpicture}  
        \def \n {8} 
        \def \radius {1.3cm} 
        \def \radiuss {2cm} 
        \def \margin {15} 

        \foreach \s in {1,...,\n}
         {
          \draw[ -, >=latex] ({360/\n * (\s -
            1)+\margin}:\radius) arc ({360/\n * (\s -
            1)+\margin}:{360/\n * (\s)-\margin}:\radius); }

        \node [draw, circle, red] at ({-360/\n * (1 - 1)+90}:\radius) {$q3$};
        \node [draw, circle, red] at ({-360/\n * (2 - 1)+90}:\radius) {$q1$};
        \node [draw, circle, red] at ({-360/\n * (3 - 1)+90}:\radius) {$q2$};
        \node [draw, circle, red] at ({-360/\n * (4 - 1)+90}:\radius) {$q4$};
        \node [draw, circle] at ({-360/\n * (5 - 1)+90}:\radius) {$5$};
        \node [draw, circle] at ({-360/\n * (6 - 1)+90}:\radius){$6$};
        \node [draw, circle] at ({-360/\n * (7 - 1)+90}:\radius) {$7$};
        \node [draw, circle] at ({-360/\n * (8 -1)+90}:\radius) {$8$}; 
        \node at ({-90}: \radiuss) {(a)};
      \end{tikzpicture}
    \hspace{0.4cm}
          \begin{tikzpicture}      
        \def \n {8} \def \radius {1.3cm} \def \radiuss {2cm} \def
        \margin {15} 

        \foreach \s in {1,...,\n} { \draw[ -, >=latex] ({360/\n * (\s -
            1)+\margin}:\radius) arc ({360/\n * (\s -
            1)+\margin}:{360/\n * (\s)-\margin}:\radius); }

        \node [draw, circle, red] at ({-360/\n * (1 - 1)+90}:\radius) {$q1$};
        \node [draw, circle, red] at ({-360/\n * (2 - 1)+90}:\radius) {$q2$};
        \node [draw, circle, red] at ({-360/\n * (3 - 1)+90}:\radius) {$q3$};
        \node [draw, circle, red] at ({-360/\n * (4 - 1)+90}:\radius) {$q4$};
        \node [draw, circle] at ({-360/\n * (5 - 1)+90}:\radius) {$5$};
        \node [draw, circle] at ({-360/\n * (6 - 1)+90}:\radius){$6$};
        \node [draw, circle] at ({-360/\n * (7 - 1)+90}:\radius) {$7$};
        \node [draw, circle] at ({-360/\n * (8 -1)+90}:\radius) {$8$}; 
        \node at ({-90}: \radiuss) {(b)};
      \end{tikzpicture}
  \end{center}
  \vspace*{-2em}
  \caption{(a) Qubit mapping to nodes if $q1$ and $q3$ swap positions. (b)
    Qubit mapping to nodes if $q2$ and $q3$ swap positions.}
  \label{fig:swap_mapping}
\end{figure}
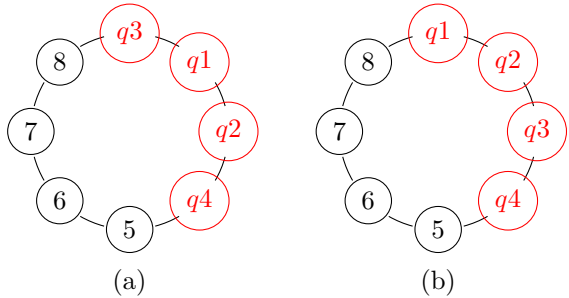

\begin{figure}[t!]
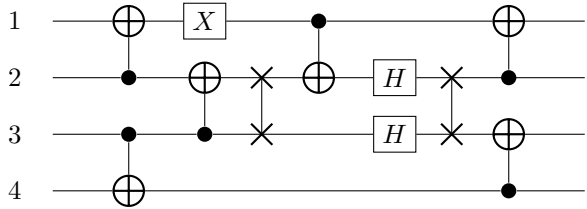

  \ctikzfig{rd-circuit-example-bis}
  \caption{Quantum circuit in Fig.~\ref{fig:circuit} mapped to
    architecture graph of Fig.~\ref{fig:complete}b.}
  \label{fig:circuit_mapped}
\end{figure}

However at the third timestep our luck has run out: the CNOT gate
between $q1$ and $q2$ is not possible in the current configuration.
We must add SWAP gates to exchange logical qubits to enable the
desired two-qubit interactions.  In the example there are two
candidates: swapping nodes 1 and 3, or swapping nodes 2 and 3,
yielding the configurations shown in Fig.~\ref{fig:swap_mapping}.
Looking ahead to the final slice -- slice 4 has no 2-qubit gates so
can be ignored -- we see that $q3$ and $q4$ will need to interact.  In
configuration (a) these qubits are distance 3 apart, and hence two
additional swaps will be needed to bring them together.  In
configuration (b) however they are already adjacent.  As we want to
minimise the number of additional elements to our circuit we choose
to swap nodes 2 and 3 to yield the final circuit shown in
Fig.~\ref{fig:circuit_mapped}.

While this was a tiny example we can see in microcosm all the key
elements of the problem: the need to find a mapping of qubits to
nodes; the notion of distance between qubits at the next timestep; and
the need to compute the permutation of the nodes to enable the next
timestep.  It should be clear even from this small example that as the
size of the circuit increases the number of candidate swaps increases
dramatically.  Further, if we have to swap several pairs of qubits at
the same time, improving the situation for one pair may worsen the
situation for another pair. There is a clear arbitrage to apply to
bring all the pairs together as soon as possible.

In the worst case $\mathcal{O}(n^2)$ swaps suffice to get from
any $n$-node configuration to any other
\cite{10.1007/978-3-319-07890-8_31}, although for sufficiently regular
graphs much better is possible \cite{brierley2015efficient}.  A recent
lower bound result states that the minimum number of swaps is $\mathcal{O}(\log
n)$ in the worst case \cite{herbert2018depth}, which is achieved by
the cyclic butterfly network \cite{brierley2015efficient}.

Our goal is to optimise the circuit globally so finding optimal mappings
between timesteps is not sufficient.  It is necessary to evaluate
candidate mappings across multiple timesteps; this is the core
of the \tket routing algorithm.




\subsection{SWAP synthesis and routing}
\label{sec:directed-graphs}

In the preceding section we described the routing problem in terms of
inserting SWAP gates into the circuit.  However not all device
technologies offer SWAP as a primitive operation. Superconducting
devices, for example, typically offer a single 2-qubit interaction
from which all other gates, including the SWAP, must be constructed.
As a further complication, these interactions may be asymmetric. For
example, in some IBM devices \cite{ibm_doc_tokyo}, the 2-qubit
interaction is a CNOT where one qubit is always the control and the
other always the target.  The graph representing the machine is
therefore directed, as shown in Fig.~\ref{fig:directed_ring}, where
the direction indicates the orientation of the gate.

\begin{figure}[h!]
  \ctikzfig{graph-ring-directed}
  \vspace*{-1em}
  \caption{Architecture with one-way connectivity constraint.}
  \label{fig:directed_ring}
\end{figure}

This complication is easily removed by the usual trick of inserting
Hadamard gates, as Fig.~\ref{fig:CNOT_flipping}. Hence the swap gate can be implemented by three (unidirectional) CNOTS
and four Hadamards, as in Fig.~\ref{fig:SWAP_flipping}.

\begin{figure}[h!]
  \ctikzfig{circuit-redirect-cnot-eqn} 
  \vspace*{-0.3cm}
  \caption{Inverting a CNOT gate for a directed graph.}
  \label{fig:CNOT_flipping}
\end{figure}

\begin{figure}[h!]
  \ctikzfig{swap-circuit-eqns} \vspace*{-0.3cm}
  \caption{Representation of a SWAP gate in terms of three consecutive
    CNOT and its inverted representation for a directed
    graph.\label{fig:SWAP_flipping}}
\end{figure}




Consider running our routed quantum circuit on the directed
architecture of Fig.~\ref{fig:directed_ring}. As this graph constrains
the direction of interactions, the quantum circuit we produced is no
longer valid.  We account for this using the inversion in
Fig.~\ref{fig:CNOT_flipping}, producing the circuit shown in
Fig.~\ref{fig:routed_circuit}.  Many simplifications are possible on
the resulting circuit, but care must be taken to ensure that the
simplified circuit is still conformant to the architecture digraph.

\begin{figure*}[ht!]
  \ctikzfig{rd-circuit-example-tris}
  \caption{Quantum circuit in Fig.~\ref{fig:circuit} routed for
    architecture graph in Fig.~\ref{fig:directed_ring}.}
  \label{fig:routed_circuit}
\end{figure*}

\section{The \tket Routing Procedure}
\label{sec:algo}

The routing algorithm implemented in \tket guarantees compilation of
any quantum circuit to any architecture, represented as simple
connected graph.  It is therefore completely hardware agnostic.  The
algorithm proceeds in four stages:  decomposing the  input circuit into
timesteps; determining an initial placement; routing across timesteps;
and a final clean-up phase.

\subsection{Slicing the circuit into timesteps}
\label{subsec:packing}

Before routing we partition the quantum circuit into timesteps.  The
circuit structure provides a natural partial ordering of the gates;
thus a greedy algorithm starting from inputs can divide the input
circuit into ``horizontal'' partitions of gates which can be executed
simultaneously.  We simply traverse the circuit adding the qubits
involved in a 2-qubit gate to the current timestep.  Since only
multiqubit interactions (such as CNOT or CZ gates) constrain the
problem, single qubit gates can be ignored\footnote{More accurately:
  while the single qubit gates can be ignored for the purposes of routing,
  they must be retained for circuit generation; for clarity we ignore
  them for now.}.  If a gate requires a
qubit already used in the previous timestep, a new timestep is
created. This procedure is repeated until all gates are assigned to a
timestep.  A timestep thus consists of a set of disjoint pairs of
(logical) qubits which represent gates scheduled for simultaneous
execution.

Applying this method to the example from Fig.~\ref{fig:circuit}
would yield the following timesteps.
\[
\begin{array}{ccl}
  1 & \mapsto & \{\;(q1,q3),(q2,q4)\;\} \\
  2 & \mapsto & \{\;(q2,q3)\;\} \\
  3 & \mapsto & \{\;(q1,q2),(q3,q4) \;\}\\
  4 & \mapsto & \{\;(q1,q2)\;\}
\end{array}
\]
Note, that this is not the same as the illustrative slicing shown in
Fig.~\ref{fig:circuit}!\\

The \emph{density} of a timestep is a measure of the number of
simultaneous gates executed.  For an $n$-qubit architecture with
single and two qubit gates, the density is
\begin{equation*}
  d=\frac{\# \text{2-qubit gates}}{\left \lfloor{\frac{n}{2}}\right \rfloor }\;.
\end{equation*}
Note that  $d=1$ where every qubit is involved in a 2-qubit gate in
this timestep;  a timestep is \emph{sparse} when its density is close
to zero. In principle, the density could be constrained to make routing
easier.  In practice this seems to make little difference, and we use
this quantity only for the analysis in Section
\ref{sec:scal-diff-hardw}.  
    
\subsection{Initial Mapping}
\label{sec:initial-mapping}

For the routing algorithm to proceed we require an initial mapping of
logical qubits (referred to as qubits) and physical qubits (referred
to as nodes).  In \tket a simple but surprisingly effective procedure
is used.

We iterate over the timesteps to construct a graph whose vertices are
qubits.  At timestep $n$ we add the edge $(q,q')$ to the graph if \textit{(i)}
this pair is present in the timestep and \textit{(ii)} both qubits $q$ and $q'$
have degree less than 2 in the current graph.  Each connected component
of the resulting graph is necessarily either a line or a ring; the
rings are broken by removing an arbitrarily chosen edge.  

Disconnected qubits in this graph correspond either to qubits which
never interact at all, or to those whose first interaction is
with a qubit whose first two interactions are with others.
These disconnected qubits are not included in the initial placement at
all; they are added later in the routing procedure.

We then select a subgraph of the architecture with high average degree
and low diameter to start from.  If the architecture is Hamiltonian
connected -- all the common architectures are\footnote{See
  Section.~\ref{sec:graph_repre_qc} and Refs.~\cite{wong1995hamilton,
    hwang2000cycles}.} -- then it is possible to map the qubit graph
to the architecture as one long line starting from a high degree
vertex within this subgraph, and greedily choosing the highest degree
available neighbour. This ensures that most of the gates in the first
two timesteps can be applied without any swaps; the only exceptions
are those gates corresponding to the edges removed when breaking
rings.

If the initial mapping cannot be completed as one long line, then the
line is split and mapped as several line segments.








\subsection{Routing}
\label{sec:routing}

The routing algorithm iteratively constructs a new circuit which
conforms to the desired architecture, taking the sliced circuit and
the current mapping of qubits to nodes as input.

The algorithm compares the current timestep of the input circuit to
the current qubit mapping.  If a gate in the current timestep requires
a qubit which has not yet been mapped, it is allocated to the nearest
available node to its partner.  All gates which can performed in the
current mapping -- all 1-qubit gates and those 2-qubit gates whose
operands are adjacent -- are immediately removed from the timestep and
added to the output circuit.  If this exhausts the current timestep,
we advance to the next; otherwise SWAPs must be added.

We define a distance vector $d(s,m)$ which approximates the number of
SWAPs needed to make timestep $s$ executable in the mapping $m$; these
vectors are ordered pointwise.  Let $s_0$ denote the current timestep,
$s_1$ for its successor, and so on, and write $\sigma \bullet m$ to
indicate the action of swap $\sigma$ upon the mapping $m$.  We compute
a sequence of sets of candidate SWAPs as follows:
\begin{align*}
  \Sigma_0 &=   \mathsf{swaps}(s_0) \\
  \Sigma_{t+1}&=  \argmin_{\sigma \in \Sigma_t} \; d(s_t, \sigma \bullet m)
\end{align*}
where $\mathsf{swaps}(s_0)$ denotes all the pertinent swaps available
at the initial timestep.  The sequence terminates either when
$\sizeof{\Sigma_t} = 1$ or after a predefined cutoff.  The selected
SWAP is added to the circuit and the mapping is updated accordingly.
We now return to the start and continue until the entire input circuit
has been consumed. 

The pointwise ordering of the distance vectors employed by \tket is
strict in the sense that $d(s,m) > d(s,\sigma\bullet m)$ implies
that for \emph{all} pairs of qubits $(q,q')$ in $s$, the longest of the shortest paths between any two paired qubits in $\sigma\bullet m$ is not longer than the longest of the shortest paths in $m$. In other words, the diameter of the subgraph composed of all pairs of qubits $(q,q')$ in $s$ should decrease strictly under the action of swap $\sigma$ on the mapping $m$.
In consequence, in some highly symmetric configurations, the algorithm
sometimes gets stuck, failing to find any candidate swap.  We employ
two strategies to overcome this.  The first is to attempt the process
again with pairs of disjoint swaps instead of individual ones.  If
this also fails then we resort to brute force: a pair of maximally
distant qubits in the current timestep are brought together using a
sequence of swaps along their shortest connecting path.  This
guarantees at least one gate may be performed, and disrupts the
symmetry of the configuration, hopefully allowing the algorithm to
escape from the bad configuration.

\noindent
\paragraph*{Remark.}
In practice there is no need to slice the circuit in advance, and in
fact better results are achieved by computing the timesteps
dynamically during routing.  The ``next slice'' is recomputed
immediately after each update of the mapping, avoiding any unnecessary
sequentialisation.

\subsection{SWAP synthesis and clean-up}
\label{sec:circ-depth-reduct}

If the target hardware does not support SWAP as a primitive operation,
then after the circuit has been routed, and the SWAPs in the routed
circuit must be replaced with hardware appropriate gates, as per
Section~\ref{sec:directed-graphs}.  While we assume that the input
circuit was already well-optimised before routing, it is usually
possible to remove some of the additional gates which are inserted
during this process in a final clean-up pass.

The essential criterion here is that any changes to the circuit must
respect the existing routing.  This can be guaranteed by using any set
of rewrite rules between 1- and 2-qubit circuits.  The routing
procedure will not insert SWAP immediately before a 2-qubit gate on
the same two qubits, but it may do so afterwards, so the possibility
to, for example, cancel consecutive CNOT gates exists.  However such
cancellation rules are the only ``true'' 2-qubit rewrites which can be
applied.  In addition, \tket uses a small
set of rewrites for fusing single qubit gates, and commuting single
qubit gates past 2-qubit gates.  The particular rewrite rules vary
according to supported gates of the hardware.




\section{Graph Representation of Quantum Computers}
\label{sec:graph_repre_qc}

We represent the architecture of a given quantum computer as a simple
connected graph, directed or undirected.  We now list some specific
architecture graphs used in this work.
   
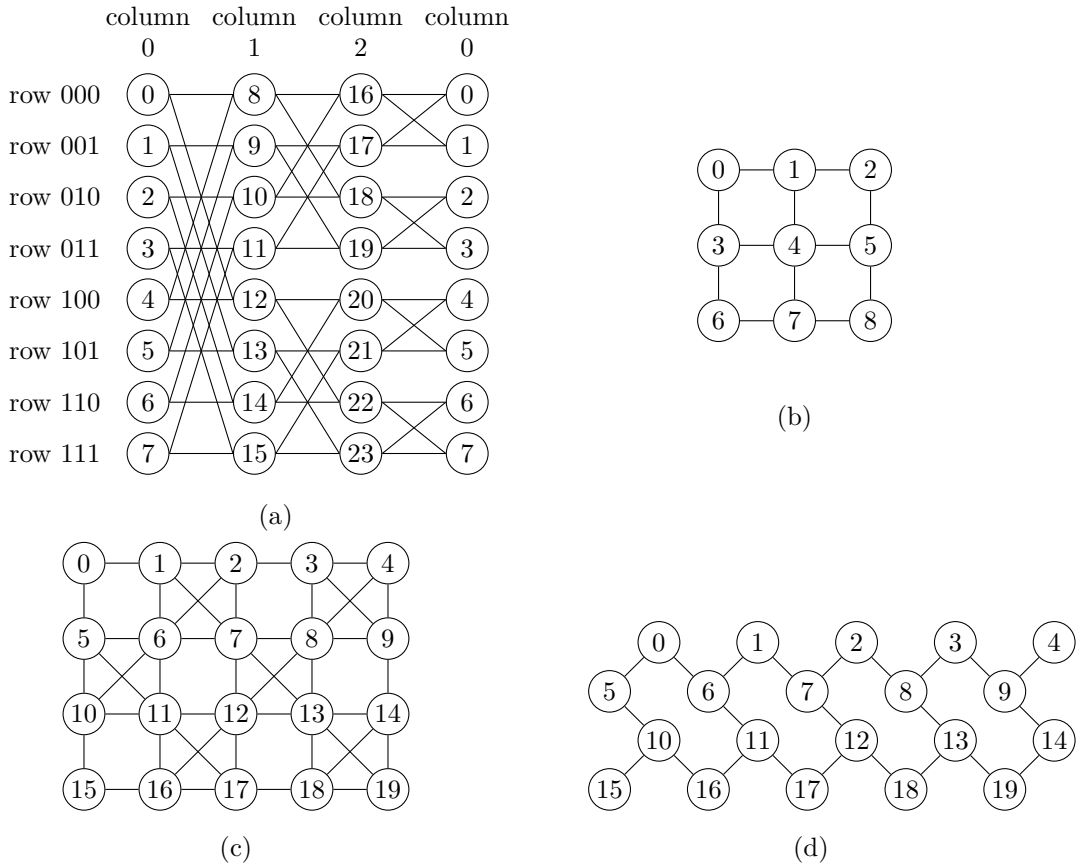
\begin{figure}[htb]
  \centering
      \begin{tikzpicture}
    \graph [left anchor=east, right anchor=west, branch down=6.8mm, grow right=14mm, nodes={draw, circle,inner sep=0pt, minimum size=0.55cm}] 
    {
      {subgraph I_n [V={0,1,2,3,4,5,6,7}, name=A] }--[butterfly={level=4}] 
      { subgraph I_n [V={8,9,10,11,12,13,14,15}, name=B] }--[butterfly={level=2}] 
     {subgraph I_n [V={16,17,18,19,20,21,22,23}, name=C] }--[butterfly={level=1}] 
      {subgraph I_n [V={0,1,2,3,4,5,6,7}, name=D] }
    };
    
     \filldraw[black] (-0.5,-0.68*0) node[anchor= east] {row 000};
      \filldraw[black] (-0.5,-0.68*1) node[anchor= east] {row 001};
       \filldraw[black] (-0.5,-0.68*2) node[anchor= east] {row 010};
        \filldraw[black] (-0.5,-0.68*3) node[anchor= east] {row 011};
         \filldraw[black] (-0.5,-0.68*4) node[anchor= east] {row 100};
          \filldraw[black] (-0.5,-0.68*5) node[anchor= east] {row 101};
           \filldraw[black] (-0.5,-0.68*6) node[anchor= east] {row 110};
            \filldraw[black] (-0.5,-0.68*7) node[anchor= east] {row 111};
            
             \filldraw[black] (0,0.38) node[anchor= south,align=center] {column \\ 0};
                      \filldraw[black] (1.4,0.38) node[anchor= south,align=center] {column \\ 1};
                               \filldraw[black] (2*1.4,0.38) node[anchor= south,align=center] {column \\ 2};
                                        \filldraw[black] (3*1.4,0.38) node[anchor= south,align=center] {column \\ 0};

    \filldraw[black] (1.2*1.4,-0.68*7.8) node[anchor= north, align=center] {(a)}; 
    \end{tikzpicture}
  \hspace{2cm}
  \raisebox{-3cm}{    \begin{tikzpicture}[darkstyle/.style={circle,draw,,inner sep=0pt, minimum size=0.55cm}]
      \foreach \x in {0,...,2}
        \foreach \y in {0,...,2} 
           {\pgfmathtruncatemacro{\label}{\x - 3 *  \y +6}
           \node [darkstyle]  (\x\y) at (\x,\y) {\label};} 
    
      \foreach \x in {0,...,2}
        \foreach \y [count=\yi] in {0,...,1}  
          \draw (\x\y)--(\x\yi) (\y\x)--(\yi\x) ;

    \filldraw[black] (1,-1) node[anchor= north, align=center] {(b)}; 
    \end{tikzpicture}}
  \hspace{2cm}

  \hspace{1cm}
      \begin{tikzpicture}[darkstyle/.style={circle,draw,,inner sep=0pt, minimum size=0.55cm}]
      \foreach \x in {0,...,4}
        \foreach \y in {0,...,3} 
           {\pgfmathtruncatemacro{\label}{\x - 5 *  \y +15}
           \node [darkstyle]  (\x\y) at (\x,\y) {\label};} 
    
      \foreach \x in {0,...,3}
        \foreach \y [count=\yi] in {0,...,2}  
          \draw (\x\y)--(\x\yi) (\y\x)--(\yi\x) ;

        \foreach \y [count=\yi] in {0,...,2}  
        {  \draw (4\y)--(4\yi) (3\y)--(4\y) ;}

       \draw (33)--(43);
       \draw (10)--(21);
          \draw (20)--(11);
             \draw (02)--(11);
                \draw (01)--(12);
                   \draw (13)--(22);
                      \draw (23)--(12);
        \draw (30)--(41);
          \draw (40)--(31);
             \draw (22)--(31);
                \draw (21)--(32);
                   \draw (33)--(42);
                      \draw (43)--(32);
    
    \filldraw[black] (2,-0.5) node[anchor= north, align=center] {(c)}; 
    
    \end{tikzpicture}
  \hspace{2cm}
      \begin{tikzpicture}[darkstyle/.style={circle,draw,inner sep=0pt, minimum size=0.55cm}]
    \def \dist {1.3cm}
    \def \distt {0.65cm}
    
      \foreach \x in {0,...,4} 
        {\pgfmathtruncatemacro{\label}{\x - 5*  0 +15}
           \node [darkstyle]  (\x0) at (\dist*\x-\dist/2,\distt*0) {\label};
           \pgfmathtruncatemacro{\label}{\x - 5*  2 +15}
           \node [darkstyle]  (\x2) at (\dist*\x-\dist/2,\distt*2) {\label};
           \pgfmathtruncatemacro{\label}{\x - 5*  1 +15}
           \node [darkstyle]  (\x1) at (\dist*\x,\distt*1) {\label};
           \pgfmathtruncatemacro{\label}{\x - 5*  3 +15}
           \node [darkstyle]  (\x3) at (\dist*\x,\distt*3) {\label};} 
    
      \foreach \x [count=\xxi]   in {0,...,3}
        { \draw (\x3)--(\x2) ;
            \draw (\x3)--(\xxi2) ;
            \draw (\x1)--(\x0) ;
            \draw (\x1)--(\xxi0) ;
             \draw (\x1)--(\x2) ;}
    
        { \draw (43)--(42) ;
        \draw (41)--(40) ;
        \draw (41)--(42) ;}
    
    \filldraw[black] (2,-0.5) node[anchor= north, align=center] {(d)}; 
    
    \end{tikzpicture}
  \caption{(a) a cyclic butterfly graph with $n=3\times 2^3$ nodes
    (the first column is represented twice to improve the readibility
    of the connectivity), (b) a 2-dimensional square grid with $n=3^2$
    nodes, (c) the IBM Q 20 Tokyo chip
    (Ref.~\cite{ibm_doc_tokyo}). and (d) the Rigetti 19Q-Acorn chip
    (Ref.~\cite{pyquil_doc_acorn}). The edges represent the allowed
    interactions between qubits. \label{fig:graph_drawing}}
\end{figure}

\begin{enumerate}
\item \textit{The ring}, Fig.~\ref{fig:complete}(b). A one-dimensional
  cyclic graph where each node is connected to its two nearest
  neighbors.
\item \textit{The cyclic butterfly},
  Fig.~\ref{fig:graph_drawing}(a). A non-planar graph with $n=r\times
  2^r$ nodes. Each node is denoted by a pair $(w,i)$ where $w$ is
  $r$-bit sequence corresponding to one of the $2^r$ rows and $i$
  represents the column. Two nodes $(w,i)$ and $(v,j)$ are connected
  if $j\equiv i+1\, [r]$ and if $w=v$ or $w$ and $v$ have only one bit
  difference at position $i$, hence the connectivity is equal to 4 for
  any node, see Ref.~\cite{brierley2015efficient}.
\item \textit{The square grid}, Fig.~\ref{fig:graph_drawing}(b). A
  two-dimensional graph with a square shape where nodes are connected
  to their four neighbors except at the edges where there can be only
  two or three neighbors.
\item \textit{The IBM Q 20 Tokyo},
  Fig.~\ref{fig:graph_drawing}(c). The graph supporting the 20-qubit
  processor produced by IBM is a two-dimensional graph with 20 nodes,
  it has a rectangle shape with some extra connectivity, see
  Ref.~\cite{ibm_doc_tokyo}.
\item \textit{The Rigetti 19Q-Acorn},
  Fig.~\ref{fig:graph_drawing}(d). The graph supporting the quantum
  processor produced by Rigetti is a two-dimensional graph with 20
  nodes, see Ref.~\cite{pyquil_doc_acorn}.
\end{enumerate}
    
In Appendix~\ref{sec:classical_sorting}
Table~\ref{table:architecture_comparison} we present the basic
properties of these graphs such as their degree and diameter, and the
depth overhead of classical sorting algorithms on these graphs.    
    
\section{Results}
\label{sec:results}

The current generation of quantum computers, the NISQ devices
\cite{preskill2018quantum}, are characterised by small numbers of
qubits and shallow circuit depths.  In this setting constant factors
are more important than asymptotic analysis, so we present two sets of
empirical results on the performance of \tket's routing algorithm.  In
the first set of results we evaluate the scaling behaviour on
synthetic inputs of increasing size.  In the second we compare the
performance of \tket against competing compiler implementations on a
set of realistic circuits.  Note that while the \tket algorithm is
very efficient, we report on the quality of the results rather than
the time or memory requirements.

\subsection{Scaling}
\label{sec:scal-diff-hardw}

The routing algorithm described in Section~\ref{sec:algo} can handle
circuits of arbitrary depth, and architectures corresponding to any
connected graph.  We now evaluate how increasing the circuit depth, and
the size and connectivity of the architecture graph influence the
depth of the routed circuit. 

As described above, routing adds SWAP gates to the circuit increasing
both its total gate count and the depth of the circuit.  Since the
total gate count depends on the particular gate set supported by the
architecture, we will consider only the increase in circuit depth here.
Therefore a reasonable figure of merit is the depth ratio: 
\begin{align*}
  R & =\frac{ \text{number of output time steps}}{ \text{ number of input time steps}}\;,
\end{align*}
where timesteps are computed as described in Section~\ref{subsec:packing}.
We define the mean depth \emph{overhead} as 
\begin{align*}
N =  \text{number of output time steps} - \text{ number of input time steps}.
\end{align*}
For a fair
comparison to classical sorting algorithms, we consider that a SWAP
gate counts as only one additional gate rather than, for example,
three when decomposed into CNOT gates, and hence will induce at most
one additional time step.

\subsubsection{Scaling with depth}
\label{sec:scaling-with-depth}

To assess the performance of \tket with respect
to increasingly deep circuits we perform the following protocol for
each of the selected architectures.
\begin{itemize}
\item We randomly generate 1000 circuits of density $d=1$ and $t$
  initial timesteps for $t \in [2,10]$.  Note that requiring $d=1$
  implies there are no single qubit gates in the circuit.
\item Use \tket to route the circuit on the chosen architecture
\item Compute $R$ for the routed circuit.
\end{itemize}

We tested using the following five architectures:
\begin{itemize}
\item a ring of size $r=64$;
\item a square grid of size $r^2=64$;
\item a cyclic butterfly of size $r2^r=64$;
\item the IBM Q 20 Tokyo ($n=20$);
\item the Rigetti 19Q-Acorn\footnote{The
    Rigetti Acorn has only 20 qubits, but due a manufacturing defect
    which only 19 are \emph{usable}. This is not
    relevant to our tests\cite{Otterbach:2017aa}.} ($n=20$).
\end{itemize}
The number of nodes for the ring, square grid and cyclic butterfly
architectures is chosen for fair comparison and similarly for the IBM
and the Rigetti ones.  To eliminate sampling bias, a single set of
64-qubit circuits was generated for the all the $n=64$ architectures,
and similarly for the $n=20$ architectures.


\begin{figure*}[t!]
\centering
\subfloat{%
\includegraphics[width=.5\linewidth]{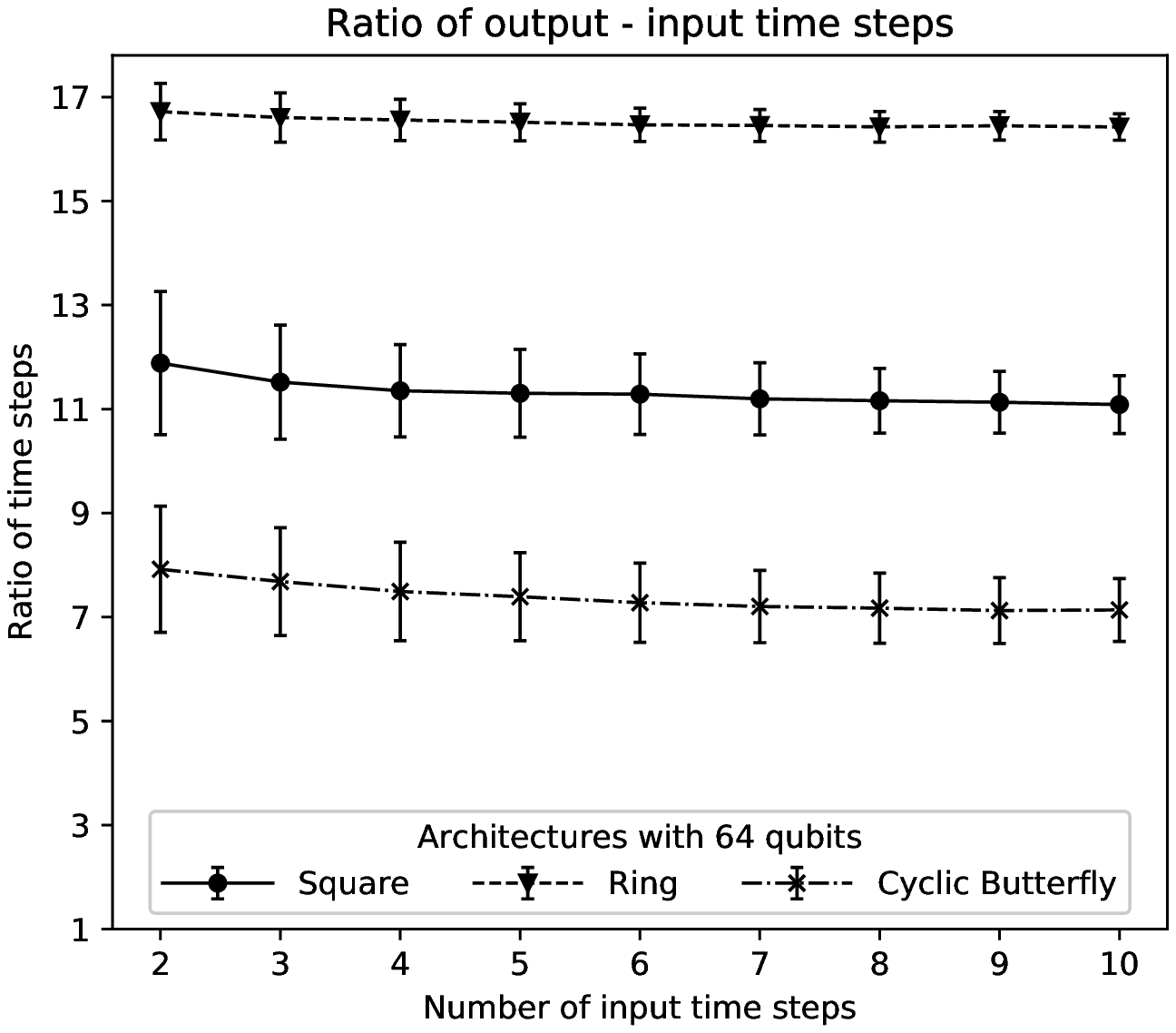}%
}\hfill
\subfloat{%
\includegraphics[width=.5\linewidth]{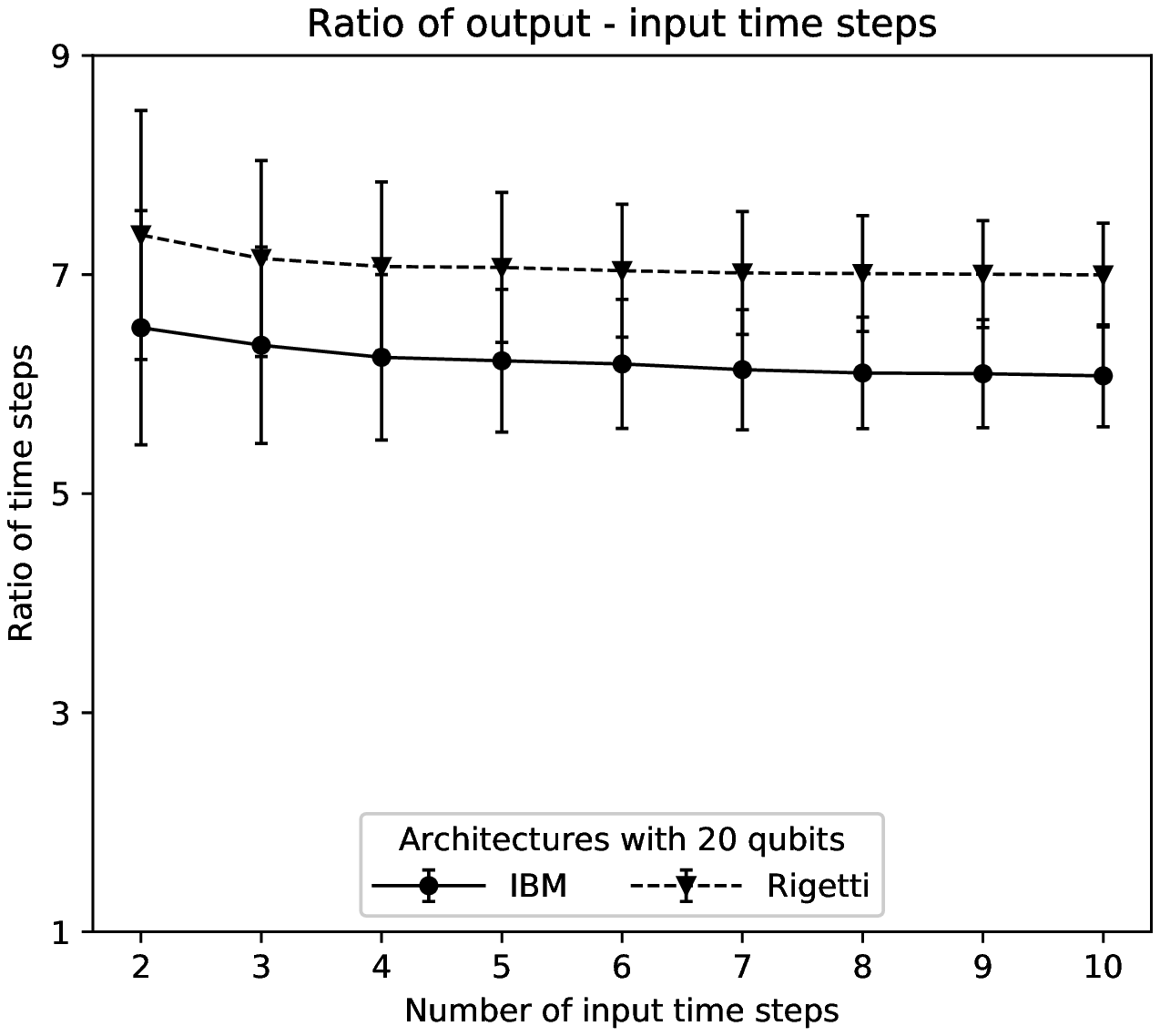}%
}
\caption{ Multiple timesteps measurement and architecture comparison. The
  mean and standard deviation of the ratio $R$ are represented. The
  left plot overlaps results for the ring, square grid and cyclic
  butterfly for 64 nodes. The right plot overlaps results for IBM and
  Rigetti architectures with 20 nodes. Results generated with random
  initial (dense) circuits with density equal to unity.}
\label{multi_timestep}
\end{figure*} 

Figure.~\ref{multi_timestep} represents the mean and standard
deviation of the ratio $R$ for the graphs.  The ratio $R$ is
approximately constant and the effect of circuit depth is dominated by
the influence of the architecture's connectivity.  This ratio seems to
converge for circuits of depth greater than 5 and we report in
Table~\ref{table:results} the values of $R$ obtained for the largest
number of input timesteps.  While the ratios obtained seem rather
large, it is worth remembering that $d=1$ circuits are the worst case
for routing.

\subsubsection{Scaling with architecture size}
\label{sec:scaling-with-graph}

To evaluate the scaling with respect to the size of the architecture
we consider single-timestep random quantum circuits of varying
density, which are routed on architectures of increasing size.
Initial qubit mapping is disabled for these tests so that only the
routing procedure is evaluated. While this is an important
part of the algorithm, in this case we are interested in the
scaling, to which the initial mapping only provides an initial offset.

\begin{itemize}
\item For each architecture of size $n$ generate $10n$ random circuits
  of depth one, for each $d \in \{0.5, 0.67, 1.0\}$.
\item Generate a random initial mapping of qubits on the
  architecture.
\item Route the timestep using \tket, using the given mapping.
\item Compute $N$ for the routed circuit.
\end{itemize}

The following architectures were evaluated:
\begin{itemize}
\item Rings of size $r \in [10,200]$
\item Square grids of size $r^2$, $r \in [3,13]$
\item Cyclic butterflies of size $r2^r$, $r \in [2,6]$.
\end{itemize}

\begin{figure*}[tb]
\makebox[\textwidth][c]{
\subfloat{%
\includegraphics[scale=0.48]{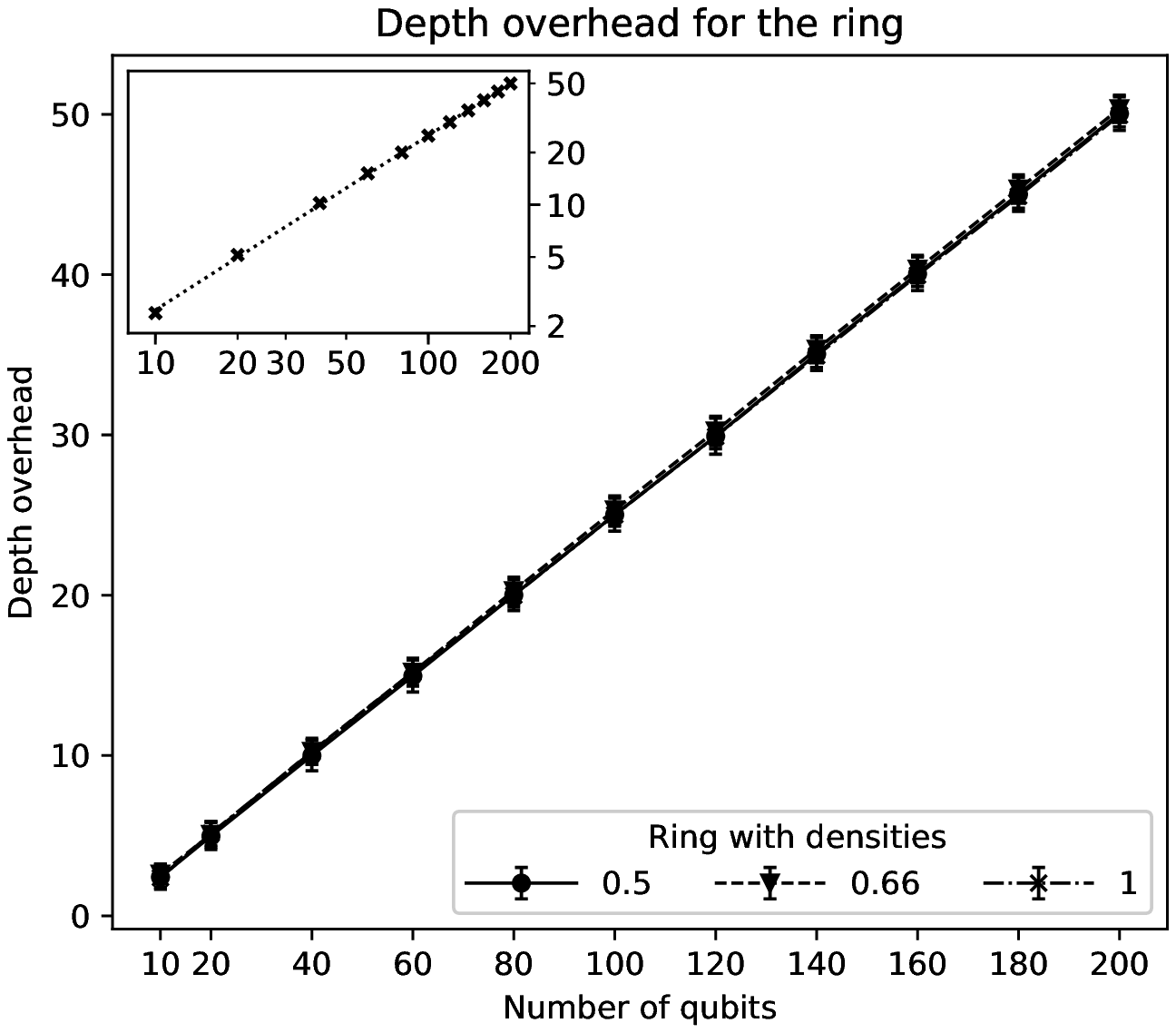}%
}\hspace*{-1.5em}
\subfloat{%
\includegraphics[scale=0.48]{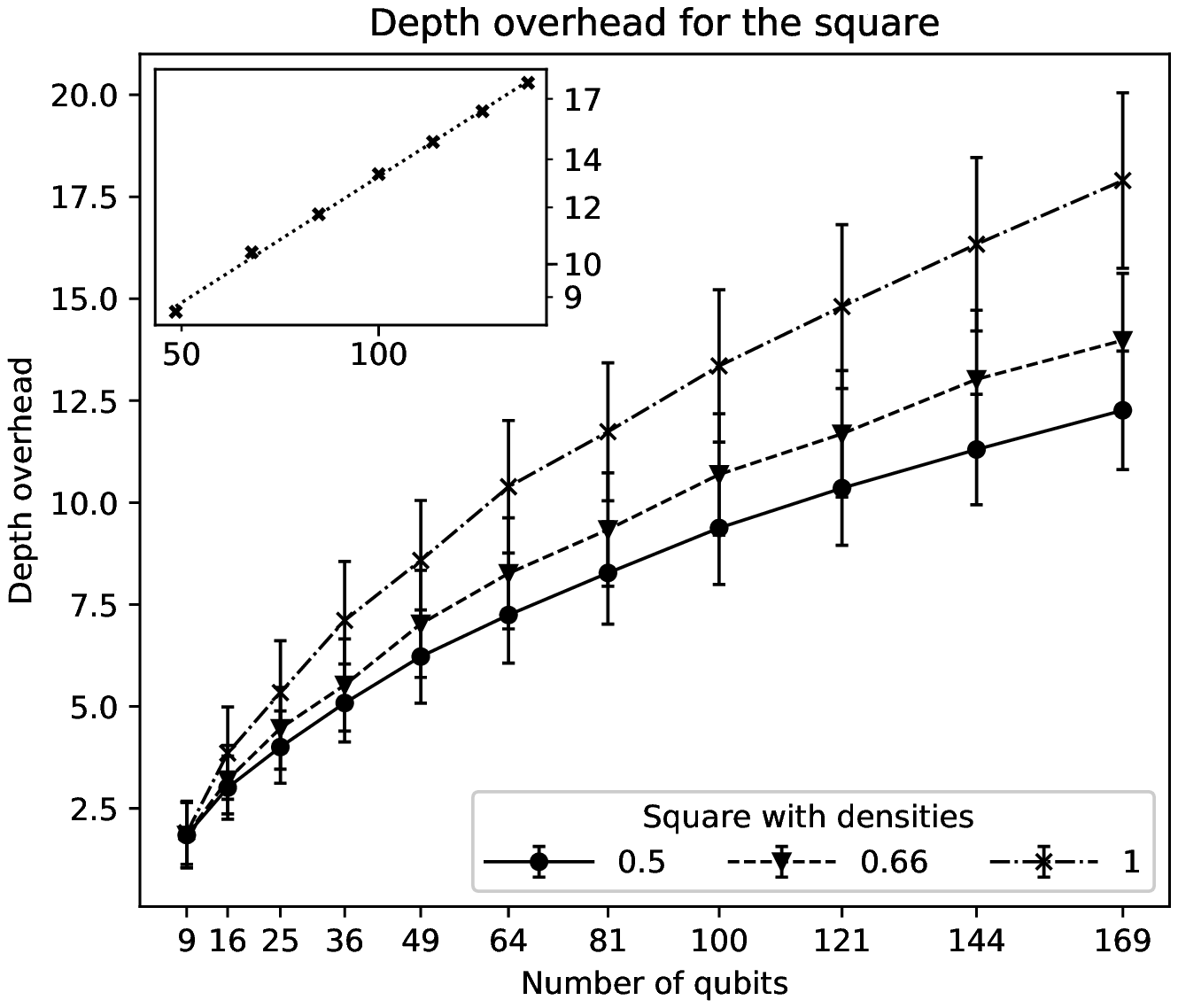}%
}\hspace*{-1.5em}
\subfloat{%
\includegraphics[scale=0.48]{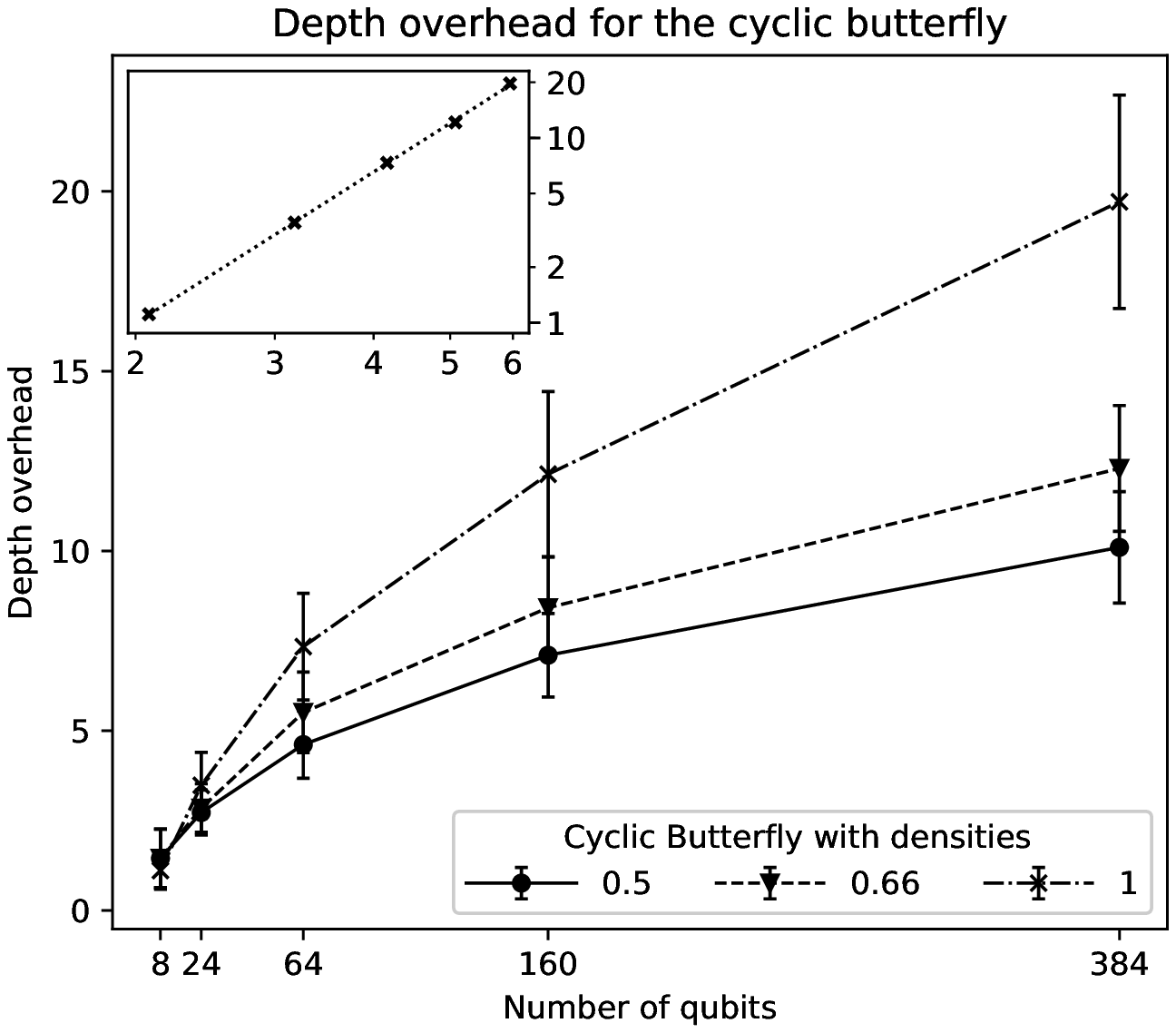}%
}
}  
\caption{Variation of depth overhead with architecture size for single
  timestep random circuits.  Plots from left to right
  the ring, square and cyclic butterfly architectures. The mean and
  standard deviation of the depth overhead versus number of nodes (or
  qubits) is represented. The inset plots represent the log-log linear
  fit for the ring and the square (resp. log-loglog fit for the
  butterfly) for the data set of density $d=1$.}
 \label{single_timestep}
\end{figure*}

The results are shown in Fig.~\ref{single_timestep} and the best fit
parameters are given in Table~\ref{tab:arch-scale}.  The prior results
for the ring and square grid are determined with a regression in
log-log space and the cyclic butterfly in log - log(log) space
(represented in the insets for $d=1$).  In each case we see that the
overhead appears to grow with the diameter of the graph, although with
an exponent that varies (slightly) with the density.


\begin{table*}
  \centering
\begin{tabular}{p{4cm} p{3.5cm} p{3.5cm} p{3.5cm}}  
  \hline

          &  &\\[-1.5ex]
    Graph & $d=0.5$ & $d=0.67$ & $d=1.0$ \\ [1ex]
    \hline 

          &  &\\[-1.5ex]
    Ring & 
    $0.2451\times  n$ & 
    $0.2451\times  n$&
    $0.2451\times  n$\\ [0.5ex]

          &  &\\[-1.5ex]
    Square & 
    $0.5501\times n^{0.55} $& 
    $0.8050\times n^{0.56}$ &
    $0.8991\times n^{0.58} $\\ [0.5ex]

          &  &\\[-1.5ex]
    Cyclic Butterfly & 
    $0.3496\times \log(n)^{1.85}$ & 
    $0.3002\times \log(n)^{2.05}$ & 
    $0.1510\times \log(n)^{2.72}$\\ [0.5ex]
    \hline
  \end{tabular}
  \caption{Scaling of the depth overhead with architecture size for single-timestep random circuits.}
  \label{tab:arch-scale}
\end{table*}

\begin{table*}
\begin{tabular}{p{3.5cm} p{7.5cm} p{5.5cm}}
   \hline
       &  &\\[-1.5ex]
   Graph & Depth overhead $N$ for single-timestep circuits & Ratio output - input timesteps $R$  \\ [1ex]
   \hline
    &  &\\[-1.5ex]
    Ring & $0.2451\times  n^{1.00}$ & $16.42 \pm 0.25 \; (n=64)$  \\ [0.5ex]

          &  &\\[-1.5ex]
       Square grid &$0.8991\times n^{0.58}$   & $11.09\pm 0.56 \; (n=64)$ \\ [0.5ex]

          &  &\\[-1.5ex]
       Cyclic butterfly & $0.1510 \times \log(n)^{2.72}$  &  $7.14\pm 0.61 \; (n=64)$ \\ [0.5ex]

          &  &\\[-1.5ex]
       Rigetti 19Q-Acorn & $\emptyset$  & $7.00\pm 0.47$  \\ [0.5ex]

      &  &\\[-1.5ex]
       IBM Q 20 Tokyo &  $\emptyset$ &  $6.08 \pm 0.47$ \\ [0.5ex]
   \hline
\end{tabular}
      \caption{Summary of our scaling results for dense circuits ($d=1$)}
         \label{table:results}
\end{table*}

\subsection{Realistic Benchmarks}
\label{sec:benchm-against-comp}

Random circuits have an essentially uniform structure, which circuits
arising from quantum algorithms typically lack.  In certain cases this
can make random circuits easier to route -- although in the preceding
section we have largely avoided this by using circuits of high
density.  To give \tket a more realistic test we have also evaluated
its performance on a standard set of 156 circuits which perform
various algorithms.  These range in size from 6 to 16 qubits, and 7 to
more than half a million gates.

\begin{figure}[t]
  \centering
\includegraphics[width=.5\linewidth]{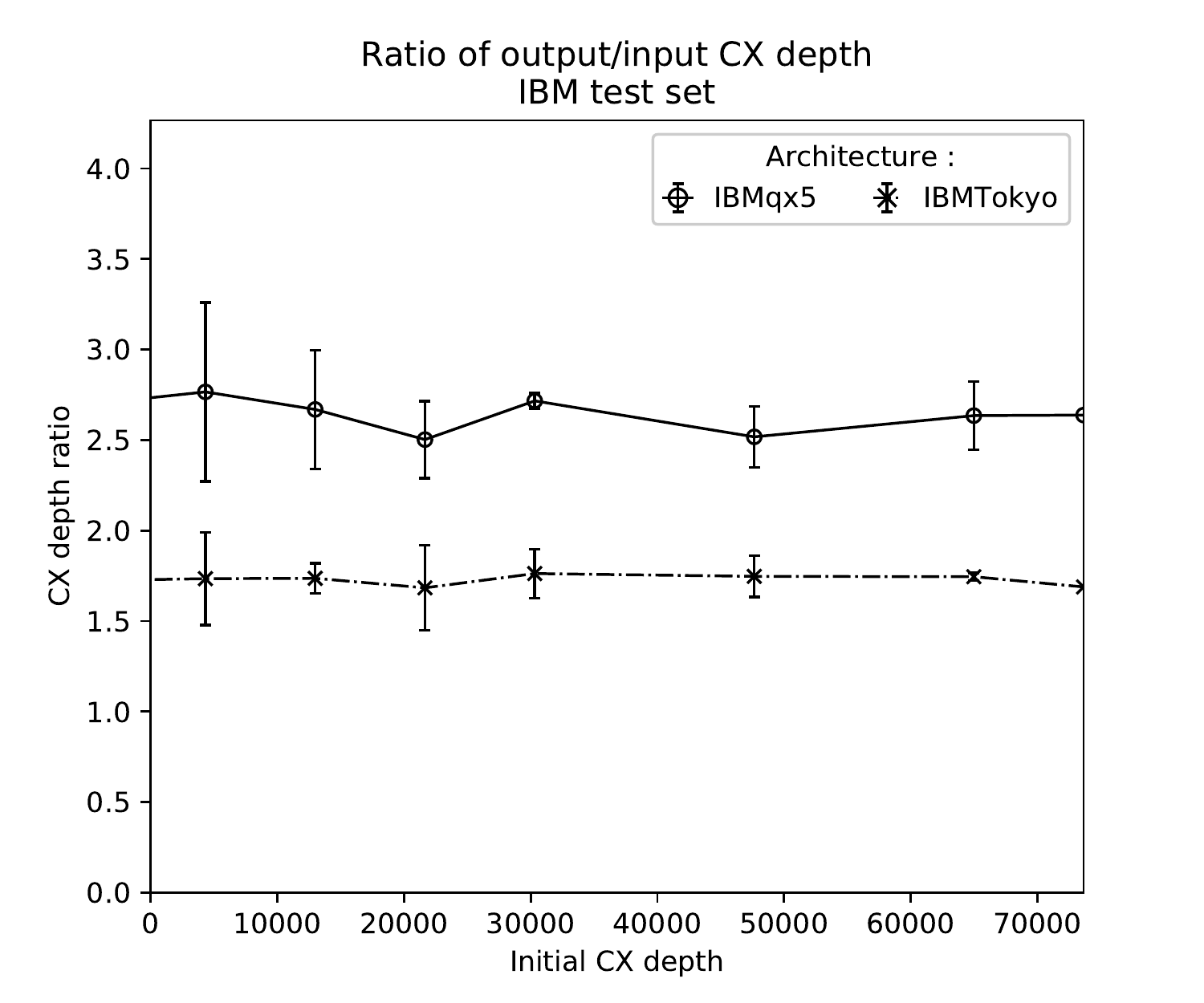}%
~\includegraphics[width=.5\linewidth]{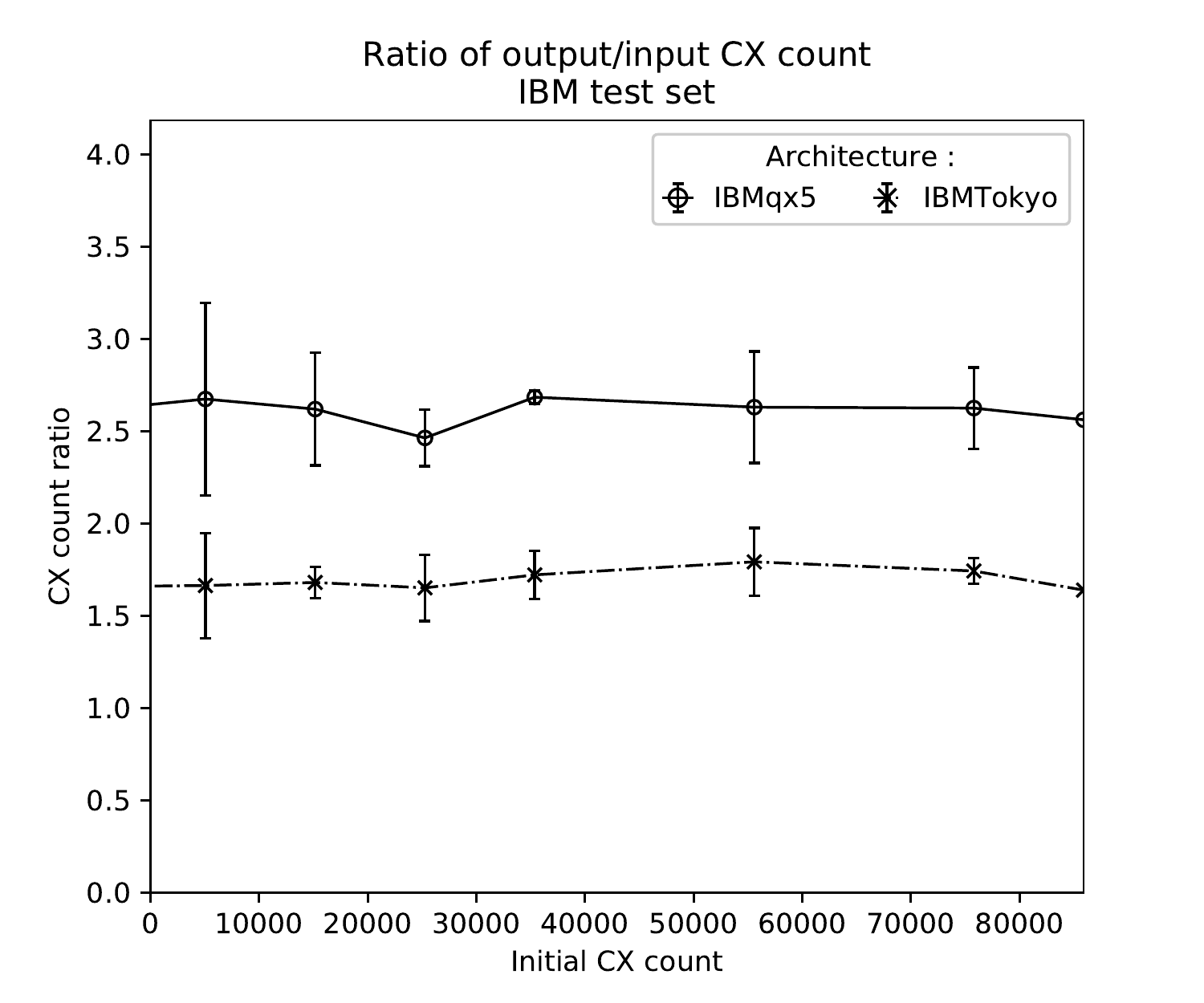}%
  
  \caption{Performance of \tket on realistic test examples.  (left) Mean ratio of output to input CX depth as a function of circuit depth (averaged in bins) (right) Mean ratio of output to input CX count (averaged in bins)}
  \label{fig:tket-cx-count-ratios}
\end{figure}

We ran \tket on each circuit of the benchmark set, with the 16-qubit
ibmqx5 \emph{Rueschlikon}, which is a $2 \times 8$ rectangular grid,
as the target architecture.  We then repeated the same test set using
the 20-qubit IBM Tokyo as the target architecture.  Since both these
architectures have CNOT as their only 2-qubit operation, and since it
has lower fidelity than the single qubit operations, we selected
figures of merit based on minimising the CNOT count and depth of the
output circuit.  In this test we do perform SWAP synthesis, to get
a more realistic evaluation of the output for these devices.  Let
$C_{CX}(c)$ be the total number of CNOT gates in circuit $c$, and let
$D_{CX}(c)$ be the depth of the circuit counting only the CNOT gates.
The two measures of interest are
\[
R_D = \frac{D_{CX}(\mathsf{out})}{D_{CX}(\mathsf{in})}
\qquad
\qquad
R_C = \frac{C_{CX}(\mathsf{out})}{C_{CX}(\mathsf{in})}
\]
where $\mathsf{in}$ and $\mathsf{out}$ are the input and output
circuits respectively.  The results are shown in Fig.~\ref{fig:tket-cx-count-ratios}.  We can see that \tket achieves
approximately linear overhead across the entire test set. The mean
$R_D$ of 2.64 and $R_C$ of $2.61$ for ibmqx5, and a mean $R_D$ of 1.73
and $R_C$ of 1.69 for IBM Tokyo.\\



\begin{figure}[tbh]
  \centering
\includegraphics[width=.5\linewidth]{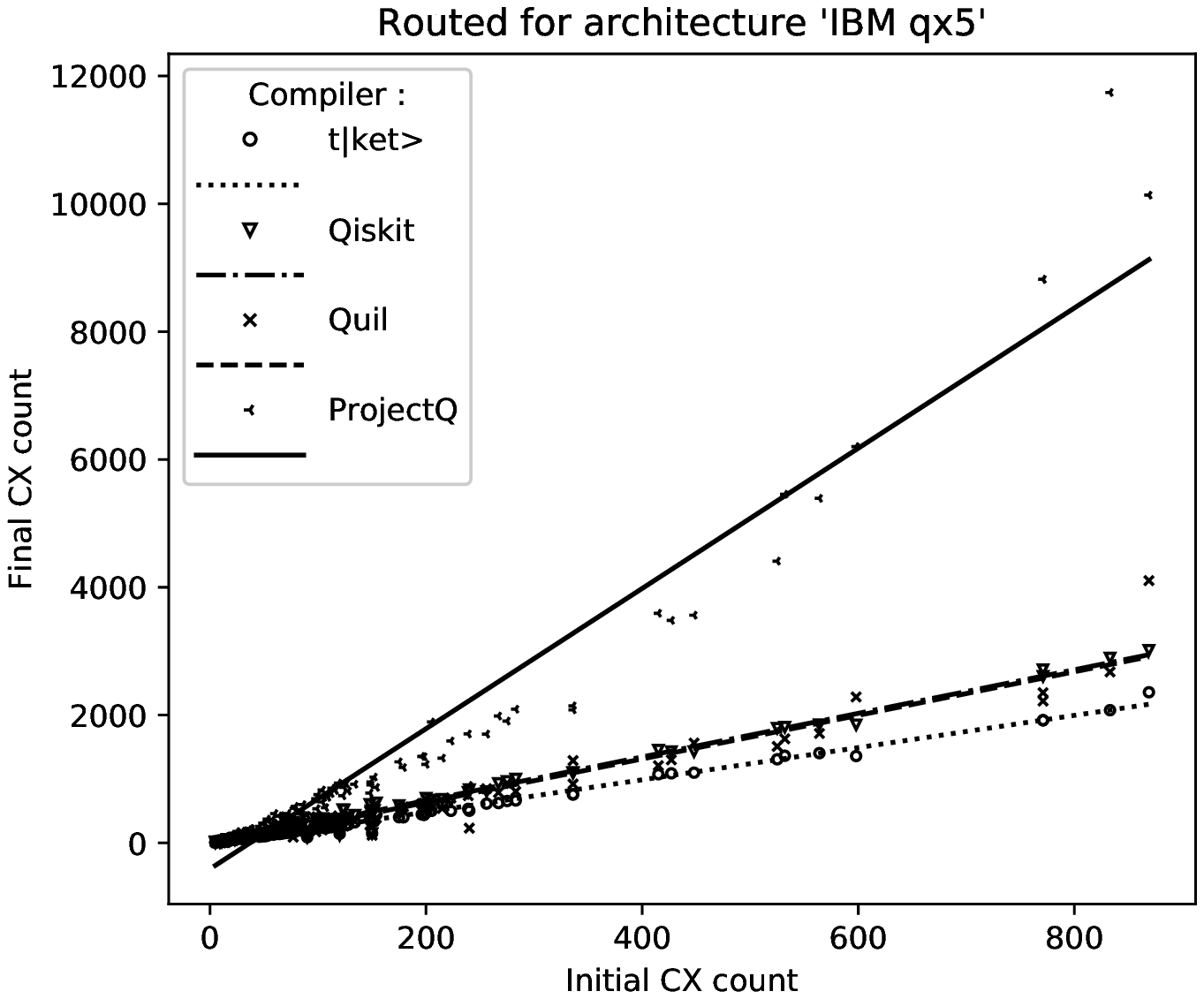}%
~\includegraphics[width=.5\linewidth]{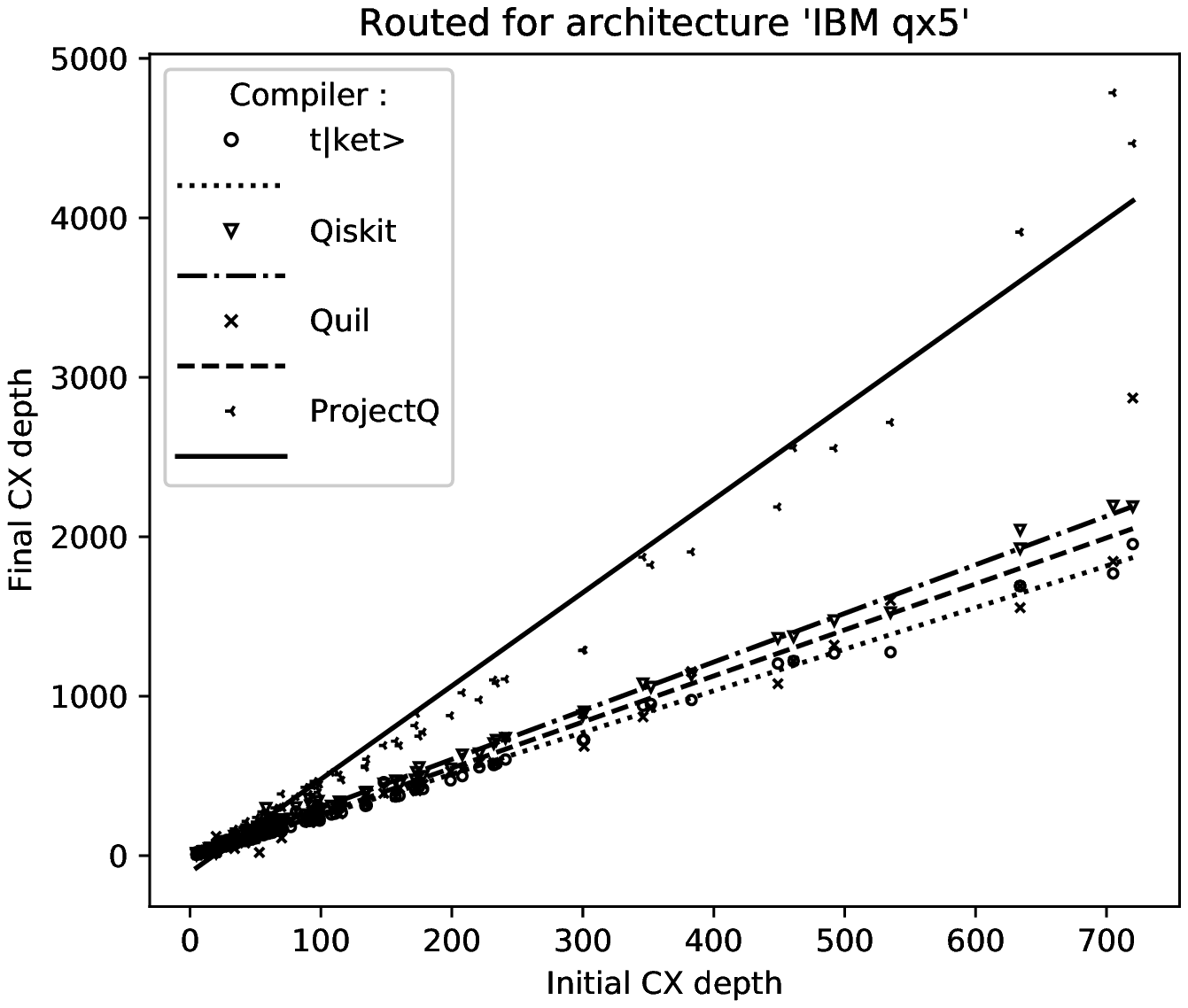}%

\includegraphics[width=.5\linewidth]{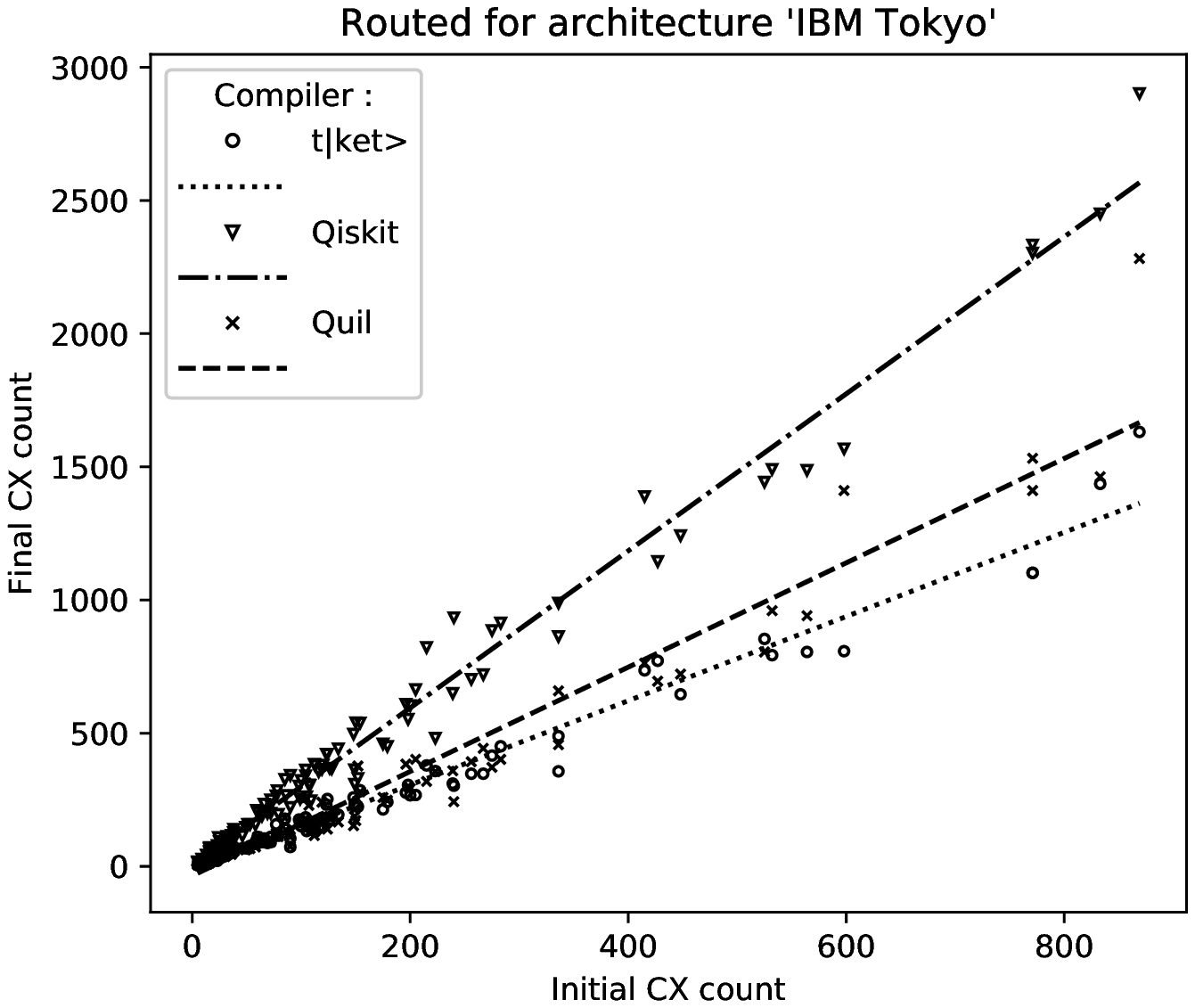}%
~\includegraphics[width=.5\linewidth]{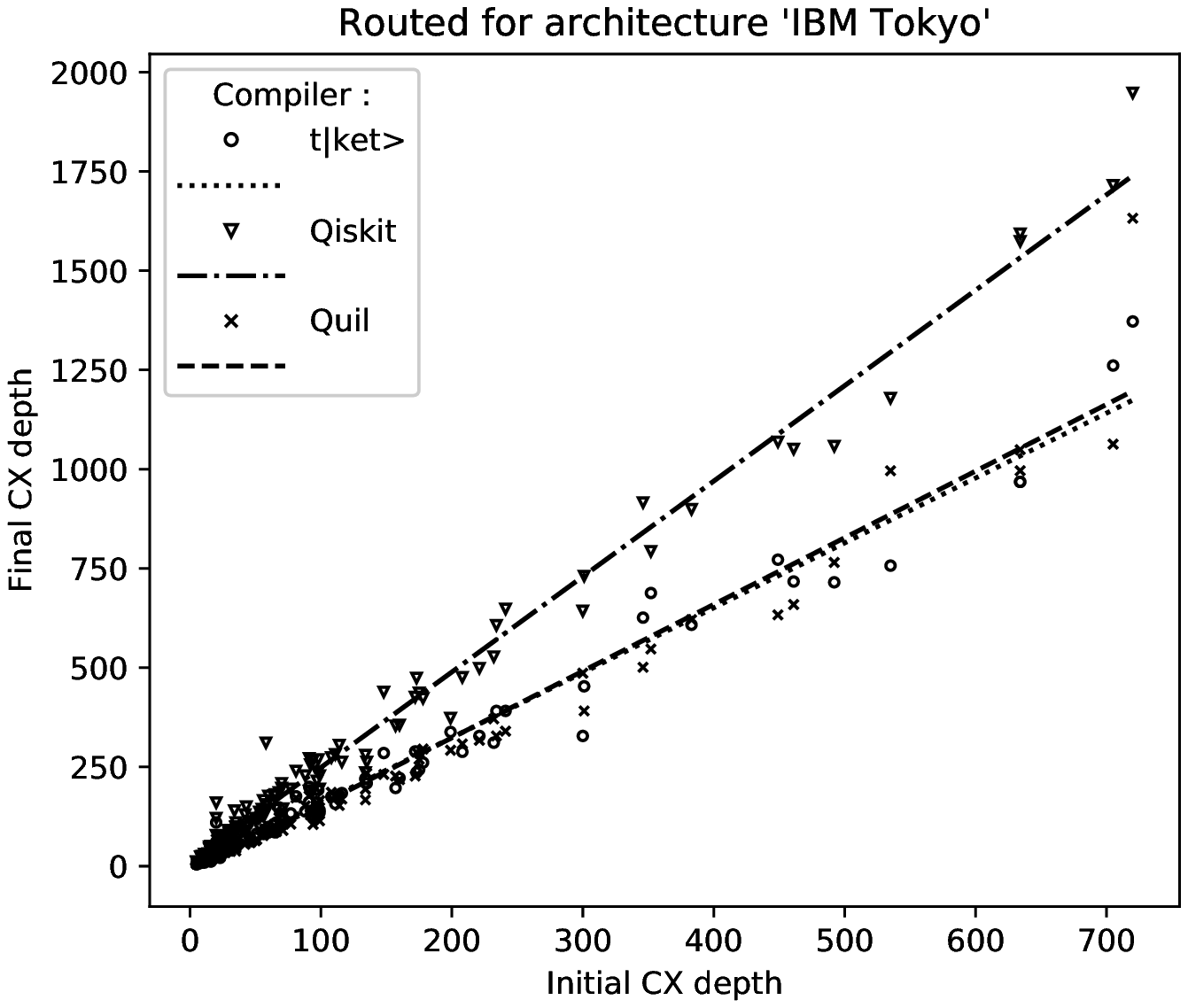}%
  
  \caption{Comparison of performance between different compilers.  Top
  row: routing on the ibmqx5 architecture.  Bottom
  row: routing on the IBM Tokyo architecture.  Left column: input CX
  count against output CX count. Right column: input CX
  depth against output CX depth. The benchmark is done against the test set available on \url{http://iic.jku.at/eda/research/ibm_qx_mapping/} and the results are averaged in bins when the initial count or depth is equal.}
  \label{fig:compare}
\end{figure}

We also compared the performance of \tket to a selection of other
freely available quantum compiler systems: IBM's QISKit \cite{qiskit},
Project Q \cite{project_q}, and Rigetti Computing's Quilc
\cite{pyquil_doc_acorn}\footnote{Since Quilc emits CZ as its preferred
2-qubit gate we computed its figures using $D_{CZ}$ and $C_{CZ}$
instead.}.  None of the other compilers was able to complete the test
set in the time allotted, despite being given at least an hour of
compute time per example on a powerful computer\footnote{See Appendix
  \ref{sec:results-tables} for more details.}.  For comparison, \tket
completed the entire benchmark set in 15 mins on the same hardware.
In addition, Project Q does not support routing for the IBM
Tokyo architecture due to its unusual graph structure; therefore it
was only tested on the ibmqx5 architecture.  Therefore comparison of
all four compilers is only available for circuits of fewer than 2000
total gates.  The comparative results are shown in Fig.~\ref{fig:compare}.  
We can see that \tket, Qiskit and Quilc exhibit
approximately linear overhead, while Project Q appears somewhat worse
than linear. A line of best fit calculated using the least squares method 
is shown for each compiler in Fig.~\ref{fig:compare}. Quilc and \tket exhibit very similar performance; the
others show significantly higher overhead.

Finally, we compared the results to the published data
of Zulehner et al. \cite{Zulehner:2017aa} who use the same benchmark
set, but use total gate count and depth as the metric.  Since Quilc
does not generate the same gate set as the others, it was excluded
from this comparison.  The algorithm of Zulehner et al.
\cite{Zulehner:2017aa} achieves comparable performance to \tket.
The results are presented in Appendix~\ref{sec:results-tables}.

\paragraph*{Where to get the test set}
\label{sec:where-get-test}

The test set we used for this work was published by IBM as part of the
QISKit Developer
Challenge\footnote{\url{https://qx-awards.mybluemix.net/\#qiskitDeveloperChallengeAward}},
a public competition to design a better routing algorithm.  The
competition was won by Zulehner et al. \cite{Zulehner:2017aa}.  The
test circuits are available from
\url{http://iic.jku.at/eda/research/ibm_qx_mapping/}. 

\section{Conclusion}
\label{sec:conclusion-outlook}

As better NISQ machines with the potential to effectively run quantum
algorithms become available, the need for software solutions that
allow users to easily run quantum circuits on them becomes more
apparent. The \tket routing module is one such solution and provides
hardware compatibility with minimal extra gate overhead. It is
flexible, general and scalable.  In this work we have outlined how the
routing procedure works and the figures of merit we use to assess
routing performance for different graphs.
   
    
Finally, we consider possible extensions of this work. Firstly, we
note that reinforcement learning offers an alternative
approach to the qubit routing problem \cite{Sherbert}. Eventually we
foresee implementing several approaches to routing in \tket to best
adapt to differing algorithms and architectures.

Secondly, when considering the routing problem, we made the implicit
assumption that all gates were equal. In real devices, notably
superconducting devices, each gates have its own fidelity and run time
and this has to be taken into account.
Splitting a quantum circuit into time steps becomes more
complex as we introduce the different run times and we also have to
ensure that the overhead in the error rate encountered by qubit is as
small as possible.  Additionally, in real life experiments, it has
been observed in \cite{tannu2018case} and
\cite{klimov2018fluctuations} that even the properties of the qubits
can fluctuate intra-days. This calls for a general protocol that could
accommodate this constraint. Addressing these different constraints
transforms the problem from a \textit{routing} one to a
\textit{scheduling} one, which we plan to address with
\tket. Implementing these constraints and measuring \tket performance
on this matter will be the object of future work.\\

    
\begin{acknowledgments} {\it Acknowledgments:}
  We thank Steven Herbert for many helpful conversations and
  encouragement.
\end{acknowledgments}

\bibliographystyle{hplain}
\bibliography{routing}

\clearpage
\appendix
   
%

\section{Dynamical routing versus sorting networks}
\label{sec:classical_sorting}

The routing problem described in this work can be solved
using classical sorting algorithms. One of these is the cyclic
odd-even sort for the ring of Fig.~\ref{fig:complete}b). Starting from
an architecture with $n$ nodes, one compares sequentially all even and
odd labeled edges.  After exactly $n-1$ time steps, the input will be
sorted regardless of input.
    
\begin{figure}[h!]
  \centerline{\includegraphics[scale=0.4]{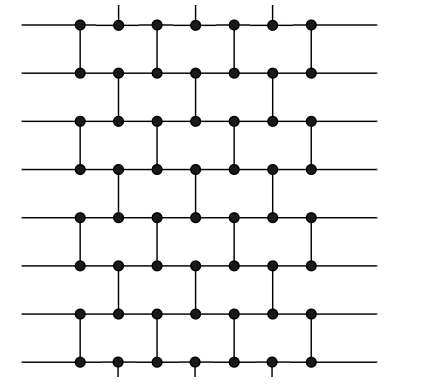}}
  \caption{An example of sorting network on 8 inputs : odd-even sort over a ring.}
  \label{oddevenextended}
\end{figure}

For the ring, square and cyclic butterfly graphs presented in
Section~\ref{sec:graph_repre_qc}, we summarize in
Table~\ref{table:architecture_comparison} some details on the degree
and diameter these graphs and the depth overhead of classical sorting
algorithms (precisely the quantity $N$ introduced in
Section~\ref{sec:results}).\\
    
The downside of classical sorting algorithms is that they are
unadapative: they compute the same sequence of comparisons regardless
of input. As circuits are usually sparse, see
Section~\ref{subsec:packing}, this leaves many unecessary comparisons,
and would treat quantum circuits as sequences of hard
timesteps. Indeed, routing solutions derived from classical sorting
algorithms tend to pack a quantum circuit into multiple timesteps and
then insert SWAP gates as in between timesteps.  Solving the routing
problem sequentially timestep by timestep produces a concatenation of
locally optimal solutions which can be very far from the globally
optimal one.  A good solution should be dynamic, consider a SWAP gates
influence on multiple timesteps, and optimize the global problem rather
than the local one.  See Ref.~\cite{zulehner2018compiling} for an additional
discussion on this matter. Additional details on sorting networks in quantum computing are
available in Ref.~\cite{brierley2,brierley2015efficient}.

\begin{table*}[ht!]
\begin{tabular}{p{5cm} p{2cm} p{3cm} p{2cm}}
   \hline
             &  &\\[-1.5ex]
 Graph &  Degree & Diameter & $N$   \\[1ex] \hline
          &  &\\[-1.5ex]
 Ring &  2 & $\frac{n-1}{2}$& $n-1$   \\ [0.5ex]

          &  &\\[-1.5ex]
 Square grid &  4 & $2\sqrt{n}-1$ & $3\sqrt{n}$  \\ [0.5ex]

          &  &\\[-1.5ex]
 Cyclic butterfly ($n=r\times 2^r)$&  4 & $\frac{3\log_2(n)}{2}$ & $6\log_2(n)$ \\ [0.5ex]

   \hline 
\end{tabular}
\caption{Comparison of different networks with $n$ nodes. }
   \label{table:architecture_comparison}
\end{table*}

\clearpage
\renewcommand{\thesubsection}{\thesection.\arabic{subsection}}
\section{Detailed Benchmark Results}
\label{sec:results-tables}

The table rows are the names of the benchmark QASM circuits, which are
available from \url{www.github.com/iic-jku/ibm_qx_mapping}. Benchmark data for Zulehner et al. is collected from results presented in
their paper \cite{Zulehner:2017aa} -- note they do not present
data for the complete set of examples.  An example Jupyter workbook
which demonstrates the benchmarking procedure is found at
\url{https://github.com/CQCL/pytket/blob/master/examples/tket_benchmarking.ipynb}.

All computations were run on a Google Cloud virtual machine with the
following specification: machine type n1-standard-2 (2 vCPUs, 7.5GB
Memory), Intel Broadwell, 16GB RAM and Standard Persistent Disk.  Each
example was run till completion, the computation aborted, or until 60
minutes of real time had passed, whichever came first.  Note that
Quilc aborts in much less than 60 minutes.

In the tables, $g$ indicates the gate count of the circuit; in
Table~\ref{tab:results-all-gates} this means all gates; in
Table~\ref{tab:results-cx-qx5} and \ref{tab:results-cx-tokyo} this
means CX count only.  The circuit depth is labelled $d$; in
Table~\ref{tab:results-all-gates} this means total depth; in
Table~\ref{tab:results-cx-qx5} and \ref{tab:results-cx-tokyo} this
means CX depth only.  The \textbf{}bold values are the best performance
on the each row.  The ``\tket comparison'' column shows the ratio
between \tket's performance and the best other compiler; values less
than 1 indicate that \tket performs better.

\paragraph*{NOTE:}
The example circuit ``ground\_state\_estimation'' gives anomalously
low values after routing.  This is due to an error in the circuit,
which allows the post-routing clean-up pass of \tket to eliminate
almost the entire circuit.

\subsection{All gates comparison on ibmqx5}
\label{sec:all-gates-comparison}

\begin{figure}[h]
  \centering
\includegraphics[width=.5\linewidth]{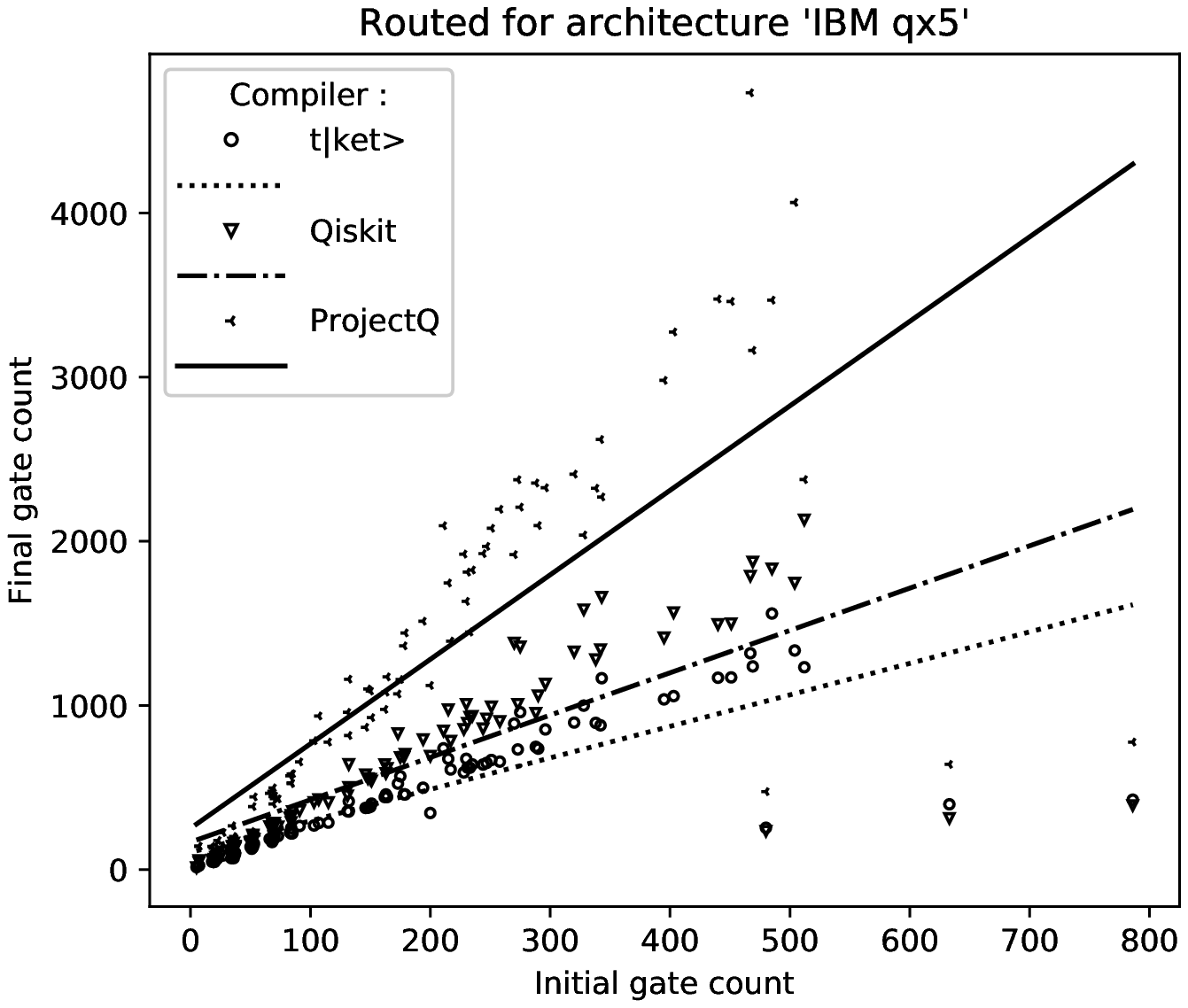}%
\includegraphics[width=.5\linewidth]{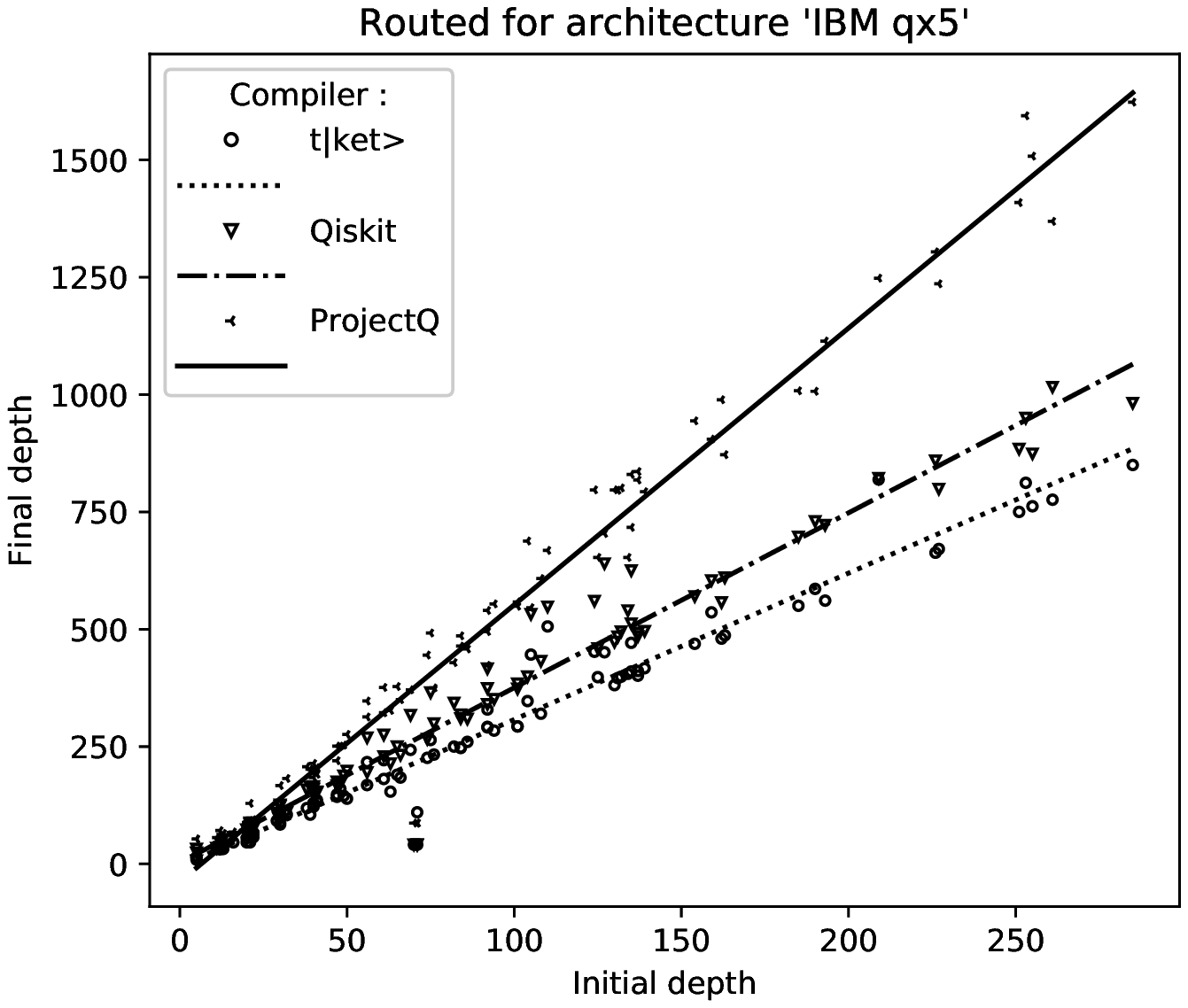}%

\includegraphics[width=.5\linewidth]{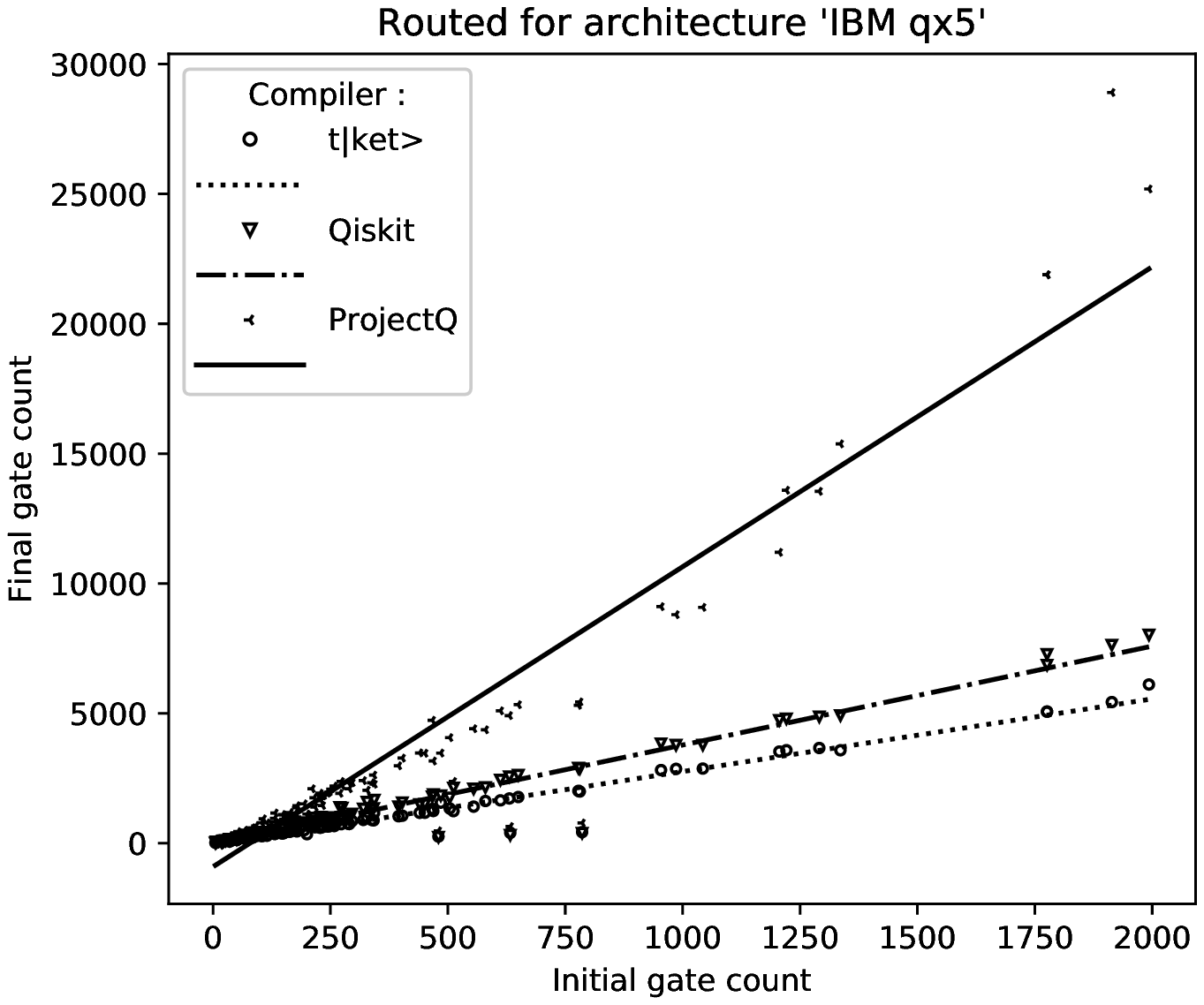}%
\includegraphics[width=.5\linewidth]{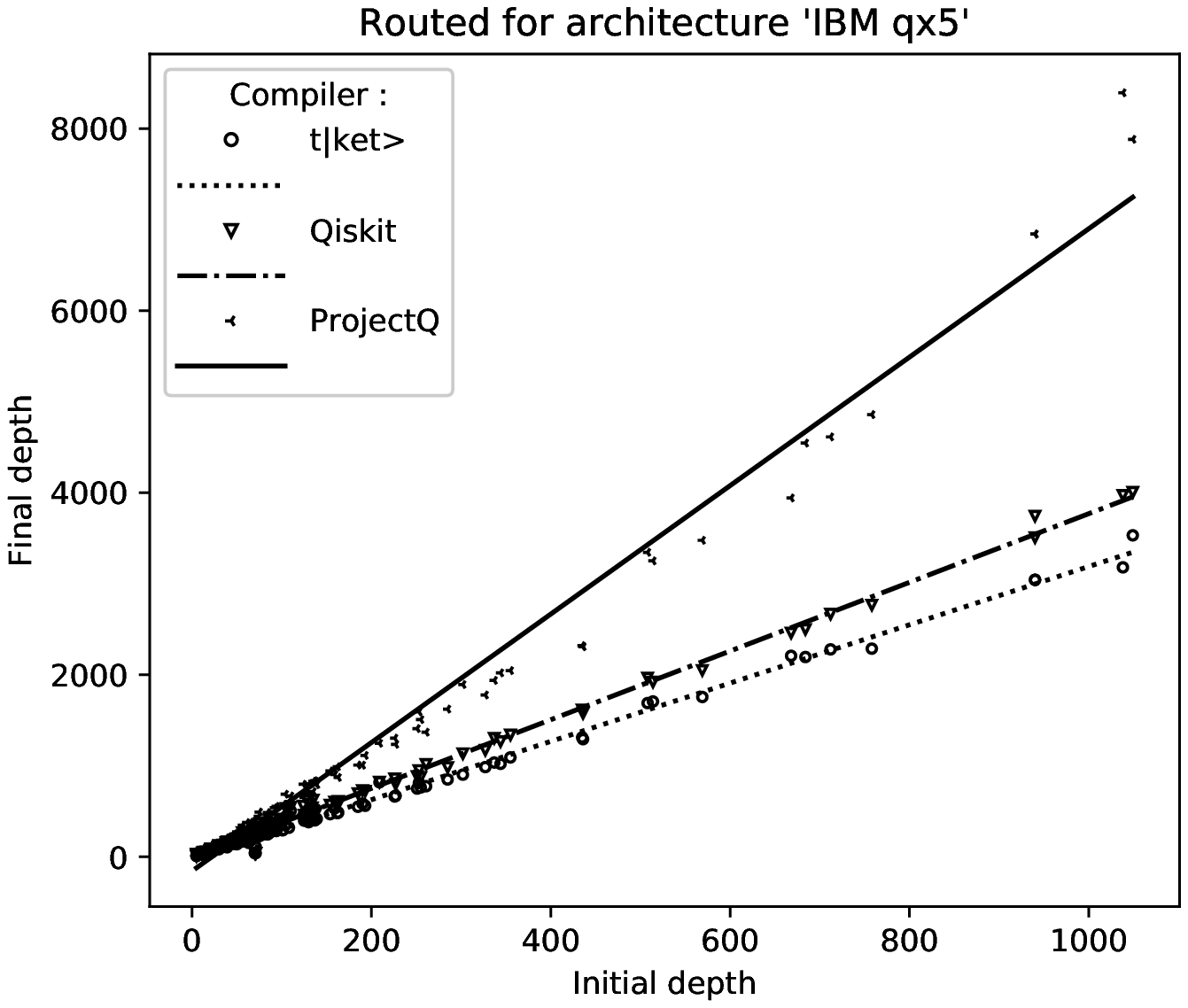}%

\caption{Routing comparison on ibmqx5, gate count and depth of the routed circuits when counting all gates. The upper
  charts are a zoomed in version of the initial segment of the lower charts. The results are averaged in bins when the initial count or depth is equal.}
\end{figure}

  {\setlength{\tabcolsep}{0.09cm}
      \begin{longtable*}{p{3.5cm}rr||rr|rr||rr|rr} 
        \caption{All gates comparison on ibmqx5}
        \label{tab:results-all-gates}\\ 
    \multicolumn{3}{c||}{}  & \multicolumn{2}{c|}{\makecell{Qiskit \\ 0.7.0}} & \multicolumn{2}{c||}{Zulehner et al.} & \multicolumn{2}{c}{CQC's \tket}&\multicolumn{2}{|c}{\makecell{\tket \\  comparison }} \\
    Name  & $g_{\rm in}$ & $d_{\rm in}$  & $g_{\rm out}$ & $d_{\rm out}$ & $g_{\rm out}$ & $d_{\rm out}$ & $g_{\rm out}$ &$d_{\rm out}$ & $r_{\rm gate}$ & $r_{\rm depth}$ \\ \hline  \endfirsthead
    
    \multicolumn{3}{c||}{}  & \multicolumn{2}{c|}{\makecell{Qiskit \\ 0.7.0}} & \multicolumn{2}{c||}{Zulehner et al.} & \multicolumn{2}{c}{CQC's \tket}&\multicolumn{2}{|c}{\makecell{\tket  \\  comparison}} \\
    Name  & $g_{\rm in}$ & $d_{\rm in}$ & $g_{\rm out}$  & $d_{\rm out}$ & $g_{\rm out}$ & $d_{\rm out}$ & $g_{\rm out}$ &$d_{\rm out}$ & $r_{\rm gate}$ & $r_{\rm depth}$ \\ \hline  \endhead    
    \hline
\endfoot    

    \hline \\
\multicolumn{11}{l}{ $g$: the number of quantum gates (elementary operations),  $d$: depth of the quantum circuits,}\\
\multicolumn{11}{l}{ -- are time-outs and * are data not provided by the Zulehner et al.}
\endlastfoot
        \begin{filecontents*}{benchmarkQX5Allgate.csv}
name,Size in,Depth in,Size_qiskit,Depth_qiskit,Size_winners,Depth_winners,Size_CQC,Depth_CQC,Size_ratioTketNextBest,Depth_ratioTketNextBest
xor5\_254,7,5,45,24,*,*,\textbf{25},\textbf{14},0.56,0.58
graycode6\_47,5,5,\textbf{13},\textbf{7},*,*,15,9,1.15,1.29
ex1\_226,7,5,56,32,*,*,\textbf{25},\textbf{14},0.45,0.44
4gt11\_84,18,11,59,34,*,*,\textbf{48},\textbf{32},0.81,0.94
4mod5-v0\_20,20,12,64,35,*,*,\textbf{50},\textbf{31},0.78,0.89
ex-1\_166,19,12,\textbf{53},36,*,*,\textbf{53},\textbf{34},1.00,0.94
4mod5-v1\_22,21,12,75,46,*,*,\textbf{64},\textbf{40},0.85,0.87
mod5d1\_63,22,13,94,53,*,*,\textbf{65},\textbf{39},0.69,0.74
ham3\_102,20,13,62,39,*,*,\textbf{47},\textbf{32},0.76,0.82
4gt11\_83,23,16,100,57,*,*,\textbf{75},\textbf{46},0.75,0.81
4gt11\_82,27,20,109,61,*,*,\textbf{86},\textbf{55},0.79,0.90
rd32-v0\_66,34,20,116,73,*,*,\textbf{70},\textbf{46},0.60,0.63
alu-v0\_27,36,21,163,86,*,*,\textbf{96},\textbf{61},0.59,0.71
4mod5-v1\_24,36,21,142,80,*,*,\textbf{97},\textbf{62},0.68,0.78
4mod5-v0\_19,35,21,143,88,*,*,\textbf{103},\textbf{66},0.72,0.75
mod5mils\_65,35,21,137,84,*,*,\textbf{93},\textbf{59},0.68,0.70
rd32-v1\_68,36,21,130,76,*,*,\textbf{70},\textbf{46},0.54,0.61
alu-v1\_28,37,22,142,84,*,*,\textbf{99},\textbf{66},0.70,0.79
alu-v2\_33,37,22,138,80,*,*,\textbf{91},\textbf{57},0.66,0.71
alu-v4\_37,37,22,130,75,*,*,\textbf{103},\textbf{64},0.79,0.85
alu-v3\_35,37,22,143,79,*,*,\textbf{103},\textbf{64},0.72,0.81
3\_17\_13,36,22,127,88,*,*,\textbf{89},\textbf{62},0.70,0.70
alu-v1\_29,37,22,135,78,*,*,\textbf{101},\textbf{66},0.75,0.85
miller\_11,50,29,158,107,*,*,\textbf{139},\textbf{92},0.88,0.86
alu-v3\_34,52,30,212,126,*,*,\textbf{146},\textbf{99},0.69,0.79
decod24-v2\_43,52,30,206,123,*,*,\textbf{136},\textbf{89},0.66,0.72
decod24-v0\_38,51,30,173,108,*,*,\textbf{127},\textbf{84},0.73,0.78
mod5d2\_64,53,32,185,110,*,*,\textbf{158},\textbf{105},0.85,0.95
4gt13\_92,66,38,263,158,*,*,\textbf{187},\textbf{119},0.71,0.75
4gt13-v1\_93,68,39,281,164,*,*,\textbf{168},\textbf{105},0.60,0.64
4mod5-v0\_18,69,40,272,160,*,*,\textbf{181},\textbf{122},0.67,0.76
decod24-bdd\_294,73,40,262,147,*,*,\textbf{205},\textbf{129},0.78,0.88
one-two-three-v2\_100,69,40,256,151,*,*,\textbf{194},\textbf{129},0.76,0.85
one-two-three-v3\_101,70,40,268,165,*,*,\textbf{199},\textbf{129},0.74,0.78
4mod5-v1\_23,69,41,283,154,*,*,\textbf{196},\textbf{136},0.69,0.88
4mod5-bdd\_287,70,41,283,152,*,*,\textbf{212},\textbf{133},0.75,0.88
rd32\_270,84,47,289,167,*,*,\textbf{220},\textbf{148},0.76,0.89
4gt5\_75,83,47,315,175,*,*,\textbf{227},\textbf{143},0.72,0.82
alu-bdd\_288,84,48,319,170,*,*,\textbf{253},\textbf{160},0.79,0.94
alu-v0\_26,84,49,331,188,*,*,\textbf{230},\textbf{145},0.69,0.77
decod24-v1\_41,85,50,355,198,*,*,\textbf{221},\textbf{139},0.62,0.70
rd53\_138,132,56,642,269,*,*,\textbf{416},\textbf{217},0.65,0.81
4gt5\_76,91,56,360,195,*,*,\textbf{266},\textbf{168},0.74,0.86
4gt13\_91,103,61,410,229,*,*,\textbf{269},\textbf{181},0.66,0.79
cnt3-5\_179,175,61,684,275,*,*,\textbf{569},\textbf{221},0.83,0.80
qft\_10,200,63,692,214,447,170,\textbf{345},\textbf{154},0.77,0.91
4gt13\_90,107,65,427,250,*,*,\textbf{284},\textbf{190},0.67,0.76
alu-v4\_36,115,66,410,232,*,*,\textbf{286},\textbf{184},0.70,0.79
mini\_alu\_305,173,69,829,317,\textbf{474},\textbf{225},524,243,1.11,1.08
ising\_model\_10,480,70,\textbf{235},\textbf{41},251,47,255,\textbf{41},1.09,1.00
ising\_model\_16,786,71,\textbf{391},\textbf{41},426,48,426,\textbf{41},1.09,1.00
ising\_model\_13,633,71,\textbf{313},\textbf{41},329,47,398,110,1.27,2.68
4gt5\_77,131,74,451,266,*,*,\textbf{356},\textbf{226},0.79,0.85
sys6-v0\_111,215,75,975,365,\textbf{613},\textbf{250},675,264,1.10,1.06
one-two-three-v1\_99,132,76,501,299,*,*,\textbf{354},\textbf{233},0.71,0.78
one-two-three-v0\_98,146,82,578,343,*,*,\textbf{376},\textbf{250},0.65,0.73
decod24-v3\_45,150,84,551,318,*,*,\textbf{383},\textbf{247},0.70,0.78
4gt10-v1\_81,148,84,555,310,*,*,\textbf{377},\textbf{248},0.68,0.80
aj-e11\_165,151,86,541,309,*,*,\textbf{402},\textbf{260},0.74,0.84
4mod7-v0\_94,162,92,641,374,*,*,\textbf{442},\textbf{291},0.69,0.78
alu-v2\_32,163,92,587,340,*,*,\textbf{458},\textbf{292},0.78,0.86
rd73\_140,230,92,1008,416,\textbf{656},\textbf{301},675,329,1.03,1.09
4mod7-v1\_96,164,94,617,351,*,*,\textbf{443},\textbf{284},0.72,0.81
4gt4-v0\_80,179,101,704,373,*,*,\textbf{458},\textbf{293},0.65,0.79
mod10\_176,178,101,685,384,*,*,\textbf{459},\textbf{293},0.67,0.76
0410184\_169,211,104,846,399,758,\textbf{336},\textbf{739},347,0.97,1.03
qft\_16,512,105,2131,532,1341,\textbf{404},\textbf{1233},446,0.92,1.10
4gt12-v0\_88,194,108,793,432,*,*,\textbf{499},\textbf{320},0.63,0.74
rd84\_142,343,110,1660,548,\textbf{971},\textbf{353},1166,506,1.20,1.43
rd53\_311,275,124,1358,560,\textbf{942},469,959,\textbf{452},1.02,0.96
4\_49\_16,217,125,783,459,*,*,\textbf{610},\textbf{398},0.78,0.87
sym9\_146,328,127,1584,640,\textbf{955},\textbf{425},999,451,1.05,1.06
4gt12-v1\_89,228,130,855,473,*,*,\textbf{592},\textbf{381},0.69,0.81
4gt12-v0\_87,247,131,919,484,*,*,\textbf{650},\textbf{396},0.71,0.82
4gt4-v0\_79,231,132,893,495,*,*,\textbf{622},\textbf{399},0.70,0.81
hwb4\_49,233,134,929,540,*,*,\textbf{621},\textbf{405},0.67,0.75
sym6\_316,270,135,1381,625,\textbf{852},\textbf{456},890,471,1.04,1.03
4gt12-v0\_86,251,135,992,512,*,*,\textbf{668},\textbf{410},0.67,0.80
4gt4-v0\_72,258,137,903,485,*,*,\textbf{658},\textbf{401},0.73,0.83
4gt4-v0\_78,235,137,935,492,*,*,\textbf{641},\textbf{411},0.69,0.84
mod10\_171,244,139,859,496,*,*,\textbf{640},\textbf{417},0.75,0.84
4gt4-v1\_74,273,154,1008,570,*,*,\textbf{732},\textbf{469},0.73,0.82
rd53\_135,296,159,1132,604,*,*,\textbf{854},\textbf{536},0.75,0.89
mini-alu\_167,288,162,953,557,*,*,\textbf{748},\textbf{480},0.78,0.86
one-two-three-v0\_97,290,163,1060,610,*,*,\textbf{737},\textbf{487},0.70,0.80
ham7\_104,320,185,1327,697,*,*,\textbf{896},\textbf{550},0.68,0.79
decod24-enable\_126,338,190,1280,730,*,*,\textbf{894},\textbf{586},0.70,0.80
mod8-10\_178,342,193,1341,722,*,*,\textbf{879},\textbf{561},0.66,0.78
cnt3-5\_180,485,209,1833,822,\textbf{1376},\textbf{669},1560,819,1.13,1.22
ex3\_229,403,226,1566,859,*,*,\textbf{1057},\textbf{663},0.67,0.77
4gt4-v0\_73,395,227,1413,799,*,*,\textbf{1037},\textbf{671},0.73,0.84
mod8-10\_177,440,251,1494,884,*,*,\textbf{1169},\textbf{750},0.78,0.85
C17\_204,467,253,1789,950,*,*,\textbf{1318},\textbf{812},0.74,0.85
alu-v2\_31,451,255,1499,874,*,*,\textbf{1171},\textbf{762},0.78,0.87
rd53\_131,469,261,1875,1016,*,*,\textbf{1238},\textbf{776},0.66,0.76
alu-v2\_30,504,285,1746,982,*,*,\textbf{1335},\textbf{850},0.76,0.87
mod5adder\_127,555,302,2094,1135,*,*,\textbf{1407},\textbf{903},0.67,0.80
rd53\_133,580,327,2149,1172,*,*,\textbf{1623},\textbf{986},0.76,0.84
cm82a\_208,650,337,2624,1303,*,*,\textbf{1777},\textbf{1036},0.68,0.80
majority\_239,612,344,2437,1267,*,*,\textbf{1648},\textbf{1022},0.68,0.81
ex2\_227,631,355,2568,1340,*,*,\textbf{1722},\textbf{1092},0.67,0.81
sf\_276,778,435,2895,1615,*,*,\textbf{2012},\textbf{1312},0.69,0.81
sf\_274,781,436,2875,1577,*,*,\textbf{1989},\textbf{1293},0.69,0.82
con1\_216,954,508,3836,1966,*,*,\textbf{2815},\textbf{1689},0.73,0.86
wim\_266,986,514,3777,1921,2985,1711,\textbf{2861},\textbf{1707},0.96,1.00
rd53\_130,1043,569,3783,2049,*,*,\textbf{2872},\textbf{1756},0.76,0.86
f2\_232,1206,668,4737,2459,*,*,\textbf{3529},\textbf{2208},0.74,0.90
cm152a\_212,1221,684,4791,2499,3738,\textbf{2155},\textbf{3580},2195,0.96,1.02
rd53\_251,1291,712,4868,2668,*,*,\textbf{3665},\textbf{2279},0.75,0.85
hwb5\_53,1336,758,4917,2766,*,*,\textbf{3577},\textbf{2287},0.73,0.83
cm42a\_207,1776,940,6862,3507,5431,\textbf{3013},\textbf{5066},3043,0.93,1.01
pm1\_249,1776,940,7274,3744,5431,\textbf{3013},\textbf{5066},3043,0.93,1.01
dc1\_220,1914,1038,7632,3973,5946,3378,\textbf{5433},\textbf{3180},0.91,0.94
squar5\_261,1993,1049,8019,4006,6267,\textbf{3448},\textbf{6110},3532,0.97,1.02
z4\_268,3073,1644,12567,6352,9717,\textbf{5335},\textbf{9379},5532,0.97,1.04
sqrt8\_260,3009,1659,12852,6615,9744,5501,\textbf{8869},\textbf{5355},0.91,0.97
radd\_250,3213,1781,13098,6880,10441,\textbf{5872},\textbf{9831},5892,0.94,1.00
adr4\_197,3439,1839,13966,7050,11301,6205,\textbf{10074},\textbf{5910},0.89,0.95
sym6\_145,3888,2187,14078,7794,*,*,\textbf{10961},\textbf{6858},0.78,0.88
misex1\_241,4813,2676,--,--,15185,\textbf{8729},\textbf{14411},8834,0.95,1.01
rd73\_252,5321,2867,--,--,*,*,\textbf{15666},\textbf{9288},,
cycle10\_2\_110,6050,3386,--,--,19857,\textbf{11141},\textbf{18361},11314,0.92,1.02
hwb6\_56,6723,3736,--,--,*,*,\textbf{18688},\textbf{11650},,
square\_root\_7,7630,3847,--,--,25212,\textbf{13205},\textbf{23258},13305,0.92,1.01
ham15\_107,8763,4819,--,--,28310,\textbf{15891},\textbf{26551},15951,0.94,1.00
dc2\_222,9462,5242,--,--,30680,\textbf{17269},\textbf{29784},18303,0.97,1.06
sqn\_258,10223,5458,--,--,32095,17801,\textbf{29740},\textbf{17456},0.93,0.98
inc\_237,10619,5863,--,--,34375,\textbf{19176},\textbf{32096},19523,0.93,1.02
cm85a\_209,11414,6374,--,--,37746,21189,\textbf{34467},\textbf{21014},0.91,0.99
rd84\_253,13658,7261,--,--,45497,24473,\textbf{41770},\textbf{24147},0.92,0.99
co14\_215,17936,8570,--,--,63826,\textbf{30366},\textbf{57906},30807,0.91,1.01
root\_255,17159,8835,--,--,57874,\textbf{30068},\textbf{52900},30448,0.91,1.01
mlp4\_245,18852,10328,--,--,*,*,\textbf{59020},\textbf{35251},,
urf2\_277,20112,11390,--,--,*,*,\textbf{64763},\textbf{37903},,
sym9\_148,21504,12087,--,--,66637,38849,\textbf{61342},\textbf{37448},0.92,0.96
life\_238,22445,12511,--,--,74632,41767,\textbf{67852},\textbf{40987},0.91,0.98
hwb7\_59,24379,13437,--,--,*,*,\textbf{68463},\textbf{41787},,
max46\_240,27126,14257,--,--,84914,\textbf{46270},\textbf{80329},46590,0.95,1.01
clip\_206,33827,17879,--,--,114336,\textbf{60882},\textbf{104857},61053,0.92,1.00
9symml\_195,34881,19235,--,--,116508,64279,\textbf{106669},\textbf{63651},0.92,0.99
sym9\_193,34881,19235,--,--,116508,64279,\textbf{106669},\textbf{63651},0.92,0.99
sao2\_257,38577,19563,--,--,131002,\textbf{66975},\textbf{120587},68114,0.92,1.02
dist\_223,38046,19694,--,--,125867,\textbf{66318},\textbf{117367},67731,0.93,1.02
urf5\_280,49829,27822,--,--,*,*,\textbf{155513},\textbf{92045},,
urf1\_278,54766,30955,--,--,*,*,\textbf{178380},\textbf{104583},,
sym10\_262,64283,35572,--,--,215569,118753,\textbf{197690},\textbf{118233},0.92,1.00
hwb8\_113,69380,38717,--,--,*,*,\textbf{201013},\textbf{122660},,
urf2\_152,80480,44100,--,--,*,*,\textbf{221274},\textbf{139339},,
urf3\_279,125362,70702,--,--,440509,239702,\textbf{406738},\textbf{237181},0.92,0.99
plus63mod4096\_163,128744,72246,--,--,439981,243861,\textbf{402395},\textbf{243814},0.91,1.00
urf5\_158,164416,89145,--,--,*,*,\textbf{465129},\textbf{284825},,
urf6\_160,171840,93645,--,--,580295,\textbf{313011},\textbf{541013},313653,0.93,1.00
urf1\_149,184864,99585,--,--,*,*,\textbf{525310},\textbf{321185},,
plus63mod8192\_164,187112,105142,--,--,640204,\textbf{354076},\textbf{585116},355168,0.91,1.00
hwb9\_119,207775,116199,--,--,655220,375105,\textbf{608175},\textbf{369246},0.93,0.98
urf3\_155,423488,229365,--,--,*,*,\textbf{1211536},\textbf{735676},,
ground\_state \_estimation\_10,390180,245614,--,--,520010,376695,\textbf{12243},\textbf{6804},0.02,0.02
urf4\_187,512064,264330,--,--,1650845,878249,\textbf{1487289},\textbf{867091},0.90,0.99
\end{filecontents*}
\csvreader[
    respect underscore, 
    late after line=\\,
    late after last line=,
    ]{benchmarkQX5Allgate.csv}
    {1=\Name, 2=\Gates, 3=\Depth, 4=\gIBM, 5=\dIBM, 6=\gWinners, 7=\dWinners, 8=\gCQC, 9=\dCQC, 10=\sTketBest, 11=\dTketBest}
    {\Name  &  \Gates & \Depth & \gIBM & \dIBM & \gWinners & \dWinners &  \gCQC & \dCQC &  \sTketBest & \dTketBest }
      \end{longtable*}
  }

\clearpage
\subsection{CX only comparison on ibmqx5}
\label{sec:cx-only-comparison}

\begin{figure}[h]
  \centering
\includegraphics[width=.5\linewidth]{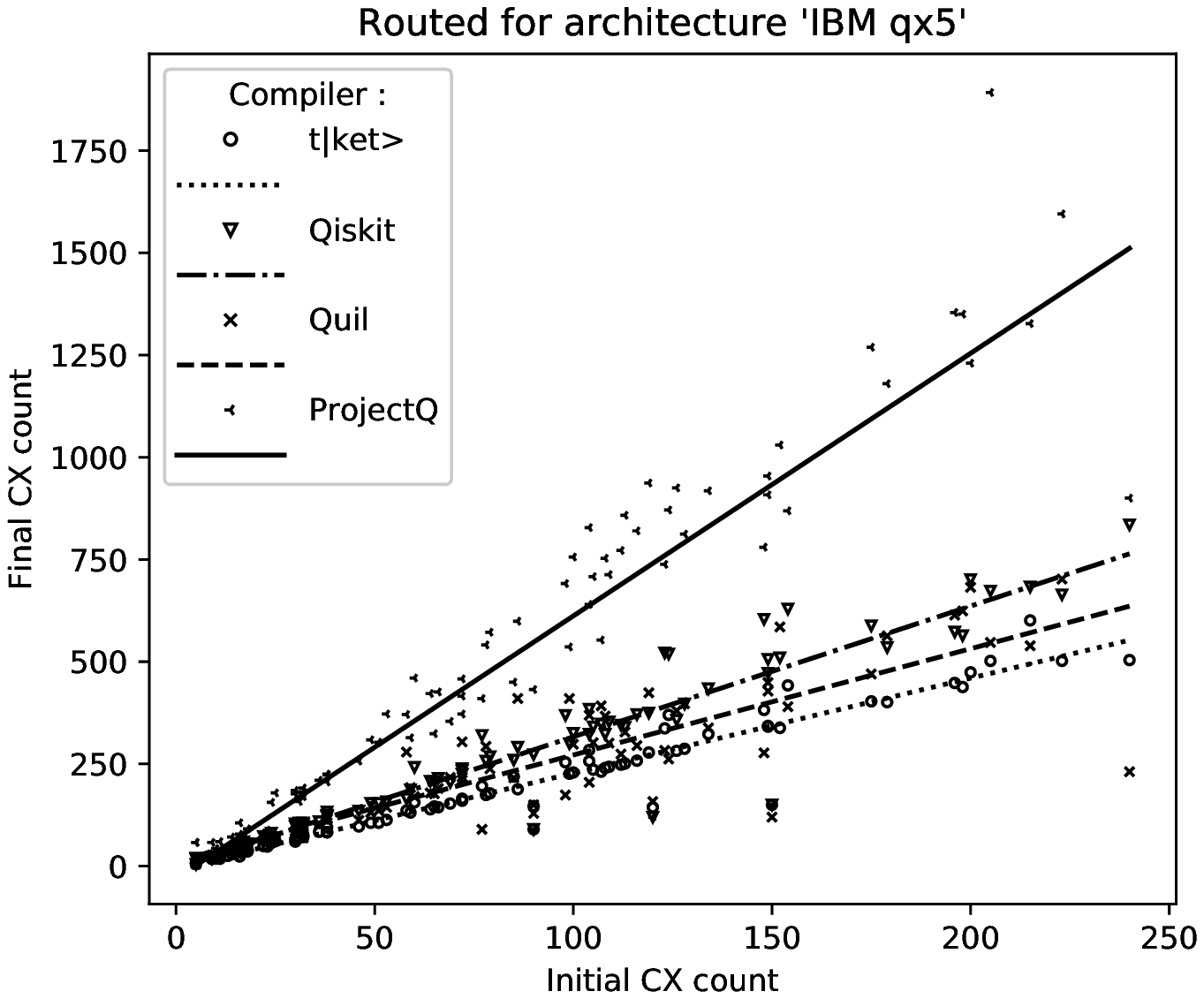}%
\includegraphics[width=.5\linewidth]{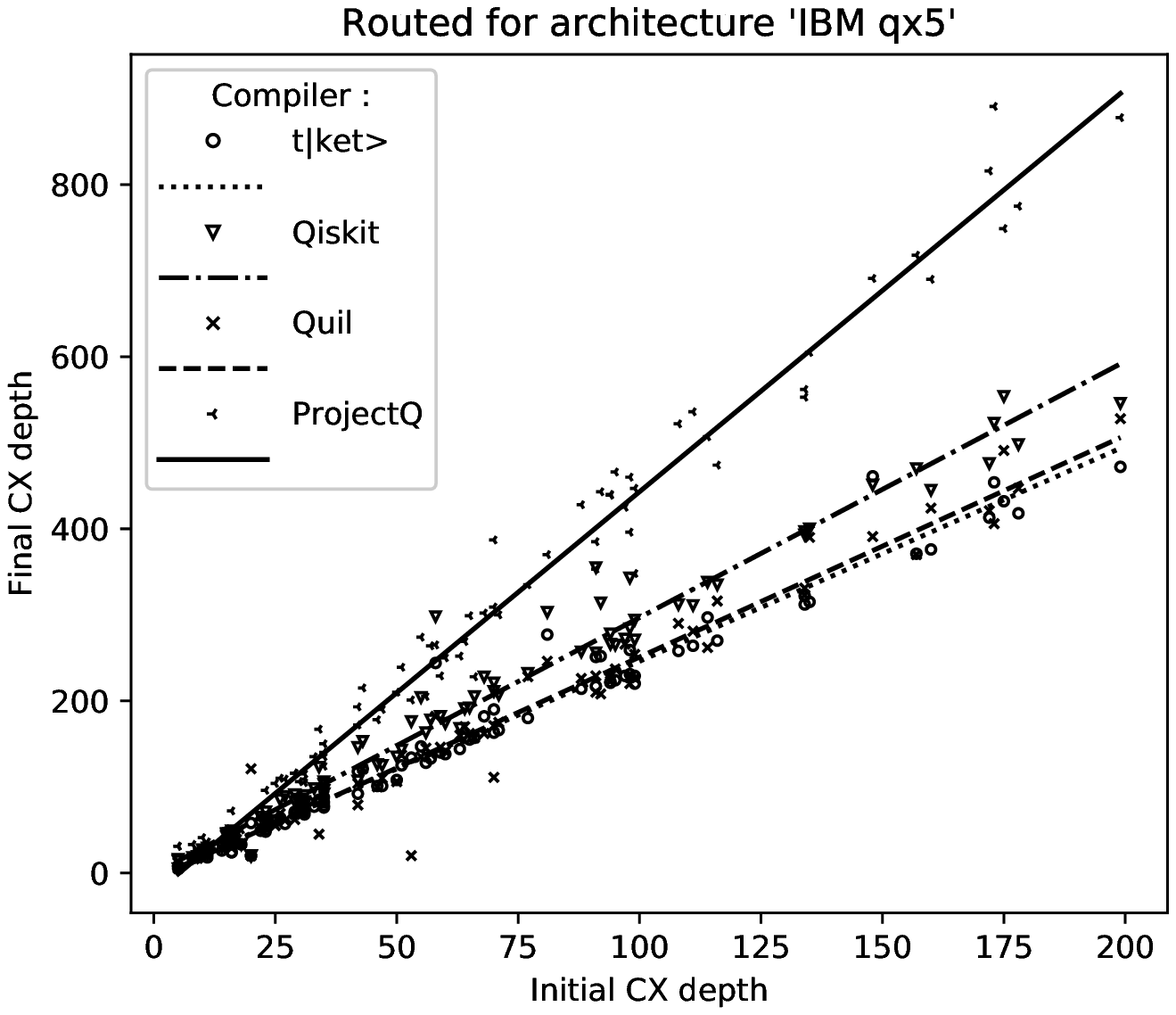}%

\caption{Routing comparison on ibmqx5, CX count and CX depth when counting only CX gates.  The  charts are a zoomed in version of the initial segment of upper charts of Fig.~\ref{fig:compare}.}
\end{figure}

  {\setlength{\tabcolsep}{0.09cm}
      \begin{longtable*}{p{3.5cm}rr||rr|rr|rr||rr|rr} 
        \caption{CX gates only comparison on ibmqx5}
        \label{tab:results-cx-qx5}\\ 
    \multicolumn{3}{c||}{}  & \multicolumn{2}{c|}{\makecell{Qiskit \\ 0.7.0}} &  \multicolumn{2}{c|}{\makecell{Project Q \\ 0.4.1}} & \multicolumn{2}{c||}{\makecell{Quilc 1.1.1 \\ Pyquil 2.1.1}}& \multicolumn{2}{c}{CQC's \tket}&\multicolumn{2}{|c}{\makecell{\tket \\  comparison }} \\
    Name  & $g_{\rm in}$ & $d_{\rm in}$ & $g_{\rm out}$ & $d_{\rm out}$ & $g_{\rm out}$ & $d_{\rm out}$ & $g_{\rm out}$ & $d_{\rm out}$ & $g_{\rm out}$ &$d_{\rm out}$ & $r_{\rm gate}$ & $r_{\rm depth}$ \\ \hline  \endfirsthead
    
    \multicolumn{3}{c||}{}  & \multicolumn{2}{c|}{\makecell{Qiskit \\ 0.7.0}}  & \multicolumn{2}{c|}{\makecell{Project Q \\ 0.4.1}} & \multicolumn{2}{c||}{\makecell{Quilc 1.1.1 \\ Pyquil 2.1.1}}& \multicolumn{2}{c}{CQC's \tket}&\multicolumn{2}{|c}{\makecell{\tket  \\  comparison}} \\
    Name  & $g_{\rm in}$ & $d_{\rm in}$ & $g_{\rm out}$ & $d_{\rm out}$ & $g_{\rm out}$ & $d_{\rm out}$ & $g_{\rm out}$ & $d_{\rm out}$ & $g_{\rm out}$ &$d_{\rm out}$ & $r_{\rm gate}$ & $r_{\rm depth}$ \\ \hline  \endhead
    
    \hline
\endfoot    

    \hline \\
\multicolumn{13}{l}{$g$: the number of quantum gates (elementary operations),}\\
\multicolumn{13}{l}{  $d$: depth of the quantum circuits and -- are time-outs}
\endlastfoot
        \begin{filecontents*}{benchmarkQX5CX.csv}
name,Size in,Depth in,Size_qiskit,Depth_qiskit,Size_projectQ,Depth_projectQ,Size_rigetti,Depth_rigetti,Size_CQC,Depth_CQC,Size_ratioTketNextBest,Depth_ratioTketNextBest
xor5\_254,5,5,17,14,58,31,17,12,\textbf{8},\textbf{8},0.47,0.67
graycode6\_47,5,5,\textbf{5},\textbf{5},\textbf{5},\textbf{5},\textbf{5},\textbf{5},\textbf{5},\textbf{5},1.00,1.00
ex1\_226,5,5,20,16,58,31,17,12,\textbf{8},\textbf{8},0.47,0.67
4gt11\_84,9,8,22,18,58,33,25,18,\textbf{18},\textbf{17},0.82,0.94
4mod5-v0\_20,10,9,22,19,43,28,20,\textbf{17},\textbf{19},18,0.95,1.06
ex-1\_166,9,9,19,19,\textbf{18},\textbf{18},27,21,\textbf{18},\textbf{18},1.00,1.00
4mod5-v1\_22,11,10,29,27,60,41,29,\textbf{21},\textbf{23},22,0.79,1.05
mod5d1\_63,13,11,37,29,48,36,40,28,\textbf{25},\textbf{22},0.68,0.79
ham3\_102,11,11,24,22,20,20,24,20,\textbf{18},\textbf{18},0.90,0.90
4gt11\_83,14,14,36,30,71,38,34,28,\textbf{29},\textbf{26},0.85,0.93
4gt11\_82,18,18,42,\textbf{33},91,50,43,34,\textbf{36},\textbf{33},0.86,1.00
rd32-v0\_66,16,16,43,40,36,31,39,32,\textbf{24},\textbf{24},0.67,0.77
alu-v0\_27,17,15,60,46,71,50,52,39,\textbf{38},\textbf{36},0.73,0.92
4mod5-v1\_24,16,15,53,44,78,51,46,38,\textbf{34},\textbf{33},0.74,0.87
4mod5-v0\_19,16,16,53,49,106,72,44,37,\textbf{37},\textbf{36},0.84,0.97
mod5mils\_65,16,16,52,47,43,41,46,36,\textbf{34},\textbf{33},0.79,0.92
rd32-v1\_68,16,16,47,40,36,31,36,29,\textbf{24},\textbf{24},0.67,0.83
alu-v1\_28,18,16,52,44,42,\textbf{36},50,40,\textbf{39},37,0.93,1.03
alu-v2\_33,17,15,53,45,72,51,52,39,\textbf{35},\textbf{33},0.67,0.85
alu-v4\_37,18,16,51,43,49,43,53,40,\textbf{39},\textbf{37},0.80,0.93
alu-v3\_35,18,16,55,45,49,43,54,40,\textbf{39},\textbf{37},0.80,0.93
3\_17\_13,17,17,47,47,41,41,50,39,\textbf{35},\textbf{35},0.85,0.90
alu-v1\_29,17,15,51,42,71,50,51,38,\textbf{38},\textbf{36},0.75,0.95
miller\_11,23,23,57,57,52,52,77,59,\textbf{50},\textbf{50},0.96,0.96
alu-v3\_34,24,23,81,71,156,96,73,62,\textbf{57},\textbf{56},0.78,0.90
decod24-v2\_43,22,22,73,65,81,67,63,53,\textbf{49},\textbf{49},0.78,0.92
decod24-v0\_38,23,23,62,56,88,65,66,52,\textbf{48},\textbf{48},0.77,0.92
mod5d2\_64,25,25,70,61,179,104,63,\textbf{55},\textbf{61},60,0.97,1.09
4gt13\_92,30,26,98,85,177,110,86,69,\textbf{66},\textbf{64},0.77,0.93
4gt13-v1\_93,30,27,104,89,185,109,79,61,\textbf{60},\textbf{57},0.76,0.93
4mod5-v0\_18,31,31,103,90,158,109,91,70,\textbf{70},\textbf{68},0.77,0.97
decod24-bdd\_294,32,31,97,79,170,105,98,77,\textbf{77},\textbf{71},0.79,0.92
one-two-three-v2\_100,32,29,96,83,191,116,\textbf{69},\textbf{62},71,71,1.03,1.15
one-two-three-v3\_101,32,29,102,91,182,116,78,\textbf{68},\textbf{72},69,0.92,1.01
4mod5-v1\_23,32,30,106,86,178,106,102,79,\textbf{74},\textbf{73},0.73,0.92
4mod5-bdd\_287,31,31,106,83,170,115,100,76,\textbf{82},\textbf{75},0.82,0.99
rd32\_270,36,35,109,92,210,124,101,\textbf{82},\textbf{84},\textbf{82},0.83,1.00
4gt5\_75,38,33,121,98,223,135,116,83,\textbf{83},\textbf{77},0.72,0.93
alu-bdd\_288,38,35,122,96,206,140,124,90,\textbf{95},\textbf{88},0.78,0.98
alu-v0\_26,38,35,125,103,224,135,115,80,\textbf{83},\textbf{78},0.72,0.97
decod24-v1\_41,38,35,133,106,225,150,118,93,\textbf{83},\textbf{76},0.70,0.82
rd53\_138,60,42,242,146,460,193,170,\textbf{104},\textbf{156},118,0.92,1.13
4gt5\_76,46,42,135,107,257,172,112,\textbf{79},\textbf{97},92,0.87,1.16
4gt13\_91,49,46,154,126,309,178,131,\textbf{100},\textbf{106},101,0.81,1.01
cnt3-5\_179,85,43,260,153,450,215,\textbf{216},\textbf{121},219,\textbf{121},1.01,1.00
qft\_10,90,34,273,123,432,167,\textbf{129},\textbf{45},147,85,1.14,1.89
4gt13\_90,53,50,158,135,372,210,144,\textbf{106},\textbf{113},108,0.78,1.02
alu-v4\_36,51,47,154,125,303,190,136,110,\textbf{106},\textbf{101},0.78,0.92
mini\_alu\_305,77,53,320,176,410,201,--,--,\textbf{196},\textbf{134},0.61,0.76
ising\_model\_10,90,20,\textbf{90},\textbf{20},\textbf{90},\textbf{20},\textbf{90},\textbf{20},\textbf{90},\textbf{20},1.00,1.00
ising\_model\_16,150,20,\textbf{150},\textbf{20},\textbf{150},\textbf{20},\textbf{150},\textbf{20},\textbf{150},\textbf{20},1.00,1.00
ising\_model\_13,120,20,\textbf{120},\textbf{20},\textbf{120},\textbf{20},\textbf{120},\textbf{20},144,58,1.20,2.90
4gt5\_77,58,51,165,143,371,239,158,\textbf{121},\textbf{136},125,0.86,1.03
sys6-v0\_111,98,55,369,204,691,274,279,\textbf{137},\textbf{254},147,0.91,1.07
one-two-three-v1\_99,59,56,187,163,314,205,174,135,\textbf{131},\textbf{128},0.75,0.95
one-two-three-v0\_98,65,59,207,182,324,229,192,145,\textbf{146},\textbf{140},0.76,0.97
decod24-v3\_45,64,57,208,178,422,264,177,146,\textbf{139},\textbf{133},0.79,0.91
4gt10-v1\_81,66,60,215,174,426,250,179,\textbf{138},\textbf{144},\textbf{138},0.80,1.00
aj-e11\_165,69,63,205,168,354,252,190,\textbf{141},\textbf{153},144,0.81,1.02
4mod7-v0\_94,72,66,239,205,372,228,217,160,\textbf{162},\textbf{157},0.75,0.98
alu-v2\_32,72,64,225,191,416,269,211,\textbf{157},\textbf{165},\textbf{157},0.78,1.00
rd73\_140,104,68,384,228,640,302,304,\textbf{170},\textbf{257},182,0.85,1.07
4mod7-v1\_96,72,65,234,192,458,299,205,162,\textbf{160},\textbf{155},0.78,0.96
4gt4-v0\_80,79,71,268,207,572,300,218,\textbf{162},\textbf{178},166,0.82,1.02
mod10\_176,78,70,257,211,541,309,238,175,\textbf{174},\textbf{163},0.73,0.93
0410184\_169,104,70,323,221,828,387,293,\textbf{171},\textbf{284},190,0.97,1.11
qft\_16,240,58,835,298,900,265,\textbf{369},\textbf{111},504,244,1.37,2.20
4gt12-v0\_88,86,77,291,232,599,335,231,182,\textbf{188},\textbf{180},0.81,0.99
rd84\_142,154,81,629,303,869,370,\textbf{410},\textbf{228},442,277,1.08,1.21
rd53\_311,124,92,519,314,871,443,390,\textbf{246},\textbf{370},252,0.95,1.02
4\_49\_16,99,91,300,256,536,353,262,\textbf{208},\textbf{226},217,0.86,1.04
sym9\_146,148,91,604,355,780,385,410,\textbf{229},\textbf{382},251,0.93,1.10
4gt12-v1\_89,100,88,325,257,756,428,277,\textbf{210},\textbf{229},214,0.83,1.02
4gt12-v0\_87,112,94,345,266,772,440,298,226,\textbf{248},\textbf{221},0.83,0.98
4gt4-v0\_79,105,94,341,278,708,439,274,223,\textbf{236},\textbf{222},0.86,1.00
hwb4\_49,107,99,348,294,553,348,302,229,\textbf{231},\textbf{220},0.76,0.96
sym6\_316,123,98,522,343,738,396,392,\textbf{254},\textbf{337},259,0.86,1.02
4gt12-v0\_86,116,98,371,284,820,460,283,\textbf{227},\textbf{258},231,0.91,1.02
4gt4-v0\_72,113,95,340,265,858,466,295,\textbf{220},\textbf{251},224,0.85,1.02
4gt4-v0\_78,109,99,353,271,713,447,328,237,\textbf{243},\textbf{229},0.74,0.97
mod10\_171,108,97,323,272,753,425,301,\textbf{226},\textbf{240},229,0.80,1.01
4gt4-v1\_74,119,108,375,312,937,522,367,266,\textbf{278},\textbf{258},0.76,0.97
rd53\_135,134,114,434,338,918,507,424,\textbf{290},\textbf{323},297,0.76,1.02
mini-alu\_167,126,111,357,311,925,536,338,\textbf{262},\textbf{282},264,0.83,1.01
one-two-three-v0\_97,128,116,397,335,812,474,380,281,\textbf{287},\textbf{270},0.76,0.96
ham7\_104,149,134,506,393,954,562,397,316,\textbf{343},\textbf{312},0.86,0.99
decod24-enable\_126,149,134,472,397,908,553,428,323,\textbf{341},\textbf{322},0.80,1.00
mod8-10\_178,152,135,510,400,1030,605,449,331,\textbf{338},\textbf{315},0.75,0.95
cnt3-5\_180,215,148,683,451,1327,691,\textbf{585},\textbf{390},601,461,1.03,1.18
ex3\_229,175,157,588,470,1269,718,539,391,\textbf{403},\textbf{371},0.75,0.95
4gt4-v0\_73,179,160,535,445,1180,690,470,\textbf{370},\textbf{401},376,0.85,1.02
mod8-10\_177,196,178,573,498,1354,775,563,424,\textbf{448},\textbf{418},0.80,0.99
C17\_204,205,173,673,523,1892,891,614,\textbf{447},\textbf{502},454,0.82,1.02
alu-v2\_31,198,172,564,476,1350,816,547,\textbf{406},\textbf{438},413,0.80,1.02
rd53\_131,200,175,701,554,1230,749,624,\textbf{421},\textbf{474},432,0.76,1.03
alu-v2\_30,223,199,664,546,1595,878,682,491,\textbf{502},\textbf{472},0.76,0.96
mod5adder\_127,239,208,793,633,1705,1022,702,528,\textbf{533},\textbf{499},0.76,0.95
rd53\_133,256,221,805,640,1703,977,739,\textbf{543},\textbf{615},554,0.83,1.02
cm82a\_208,283,234,1003,724,2092,1081,848,581,\textbf{665},\textbf{576},0.78,0.99
majority\_239,267,232,932,704,1985,1102,805,581,\textbf{624},\textbf{568},0.78,0.98
ex2\_227,275,241,965,737,1907,1107,789,\textbf{584},\textbf{656},603,0.83,1.03
sf\_276,336,301,1104,902,2083,1290,935,743,\textbf{762},\textbf{727},0.81,0.98
sf\_274,336,300,1095,878,2145,1287,922,\textbf{686},\textbf{757},725,0.82,1.06
con1\_216,415,346,1454,1078,3593,1872,1288,\textbf{891},\textbf{1075},940,0.83,1.05
wim\_266,427,352,1432,1058,3482,1823,1209,\textbf{870},\textbf{1089},951,0.90,1.09
rd53\_130,448,383,1430,1132,3563,1905,1300,\textbf{930},\textbf{1099},976,0.85,1.05
f2\_232,525,449,1796,1364,4409,2187,1563,\textbf{1154},\textbf{1309},1205,0.84,1.04
cm152a\_212,532,461,1813,1374,5455,2557,1510,\textbf{1078},\textbf{1364},1221,0.90,1.13
rd53\_251,564,492,1851,1474,5391,2555,1629,\textbf{1221},\textbf{1405},1268,0.86,1.04
hwb5\_53,598,535,1850,1525,6201,2717,1713,1320,\textbf{1360},\textbf{1276},0.79,0.97
cm42a\_207,771,634,2607,1926,8818,3910,2284,\textbf{1600},\textbf{1920},1691,0.84,1.06
pm1\_249,771,634,2713,2043,8818,3910,2220,\textbf{1555},\textbf{1920},1691,0.86,1.09
dc1\_220,833,705,2895,2194,11741,4784,2351,\textbf{1690},\textbf{2077},1770,0.88,1.05
squar5\_261,869,720,3015,2189,10135,4466,2676,\textbf{1845},\textbf{2358},1954,0.88,1.06
z4\_268,1343,1112,4744,3473,22664,8032,4104,\textbf{2870},\textbf{3583},3059,0.87,1.07
sqrt8\_260,1314,1121,4853,3633,22130,8209,4107,\textbf{2937},\textbf{3416},2985,0.83,1.02
radd\_250,1405,1210,4952,3782,23920,8909,4290,\textbf{3027},\textbf{3768},3281,0.88,1.08
adr4\_197,1498,1249,5284,3883,22303,8649,4538,\textbf{3142},\textbf{3871},3279,0.85,1.04
sym6\_145,1701,1499,5356,4324,40194,11879,4910,\textbf{3675},\textbf{4170},3820,0.85,1.04
misex1\_241,2100,1797,--,--,20756,10450,5798,\textbf{4367},\textbf{5578},4941,0.96,1.13
rd73\_252,2319,1963,--,--,45433,15172,--,--,\textbf{6007},\textbf{5173},0.13,0.34
cycle10\_2\_110,2648,2276,--,--,50777,17976,--,--,\textbf{7084},\textbf{6316},0.14,0.35
hwb6\_56,2952,2559,--,--,72955,21358,--,--,\textbf{7125},\textbf{6496},0.10,0.30
square\_root\_7,3089,2520,--,--,37275,16316,--,--,\textbf{8928},\textbf{7401},0.24,0.45
ham15\_107,3858,3273,--,--,46860,21421,--,--,\textbf{10206},\textbf{8884},0.22,0.41
dc2\_222,4131,3518,--,--,45200,21119,--,--,\textbf{11545},\textbf{10210},0.26,0.48
sqn\_258,4459,3719,--,--,93394,29983,--,--,\textbf{11425},\textbf{9743},0.12,0.32
inc\_237,4636,3928,--,--,39432,21180,--,--,\textbf{12381},\textbf{10872},0.31,0.51
cm85a\_209,4986,4256,--,--,71744,29352,--,--,\textbf{13297},\textbf{11722},0.19,0.40
rd84\_253,5960,4917,--,--,118473,40259,--,--,\textbf{16027},\textbf{13432},0.14,0.33
co14\_215,7840,5759,--,--,84431,36897,--,--,\textbf{22357},\textbf{17144},0.26,0.46
root\_255,7493,5965,--,--,128324,47123,--,--,\textbf{20332},\textbf{16951},0.16,0.36
mlp4\_245,8232,6930,--,--,50107,31076,--,--,\textbf{22759},\textbf{19640},0.45,0.63
urf2\_277,10066,8312,--,--,243218,71917,--,--,\textbf{25127},\textbf{21186},0.10,0.29
sym9\_148,9408,8062,--,--,241725,73636,--,--,\textbf{23489},\textbf{20883},0.10,0.28
life\_238,9800,8356,--,--,221220,72011,--,--,\textbf{26064},\textbf{22788},0.12,0.32
hwb7\_59,10681,9112,--,--,264057,78857,--,--,\textbf{26033},\textbf{23163},0.10,0.29
max46\_240,11844,9657,--,--,269243,84007,--,--,\textbf{30709},\textbf{25891},0.11,0.31
clip\_206,14772,12028,--,--,213901,85381,--,--,\textbf{40504},\textbf{34045},0.19,0.40
9symml\_195,15232,12849,--,--,--,--,--,--,\textbf{40897},\textbf{35349},,
sym9\_193,15232,12849,--,--,--,--,--,--,\textbf{40897},\textbf{35349},,
sao2\_257,16864,13209,--,--,--,--,--,--,\textbf{46536},\textbf{37963},,
dist\_223,16624,13274,--,--,--,--,--,--,\textbf{45230},\textbf{37762},,
urf5\_280,23764,19888,--,--,--,--,--,--,\textbf{60233},\textbf{51479},,
urf1\_278,26692,22307,--,--,--,--,--,--,\textbf{69370},\textbf{58597},,
sym10\_262,28084,23736,--,--,--,--,--,--,\textbf{76180},\textbf{65905},,
hwb8\_113,30372,26041,--,--,--,--,--,--,\textbf{77247},\textbf{68560},,
urf2\_152,35210,32247,--,--,--,--,--,--,\textbf{84345},\textbf{77320},,
urf3\_279,60380,50568,--,--,--,--,--,--,\textbf{158115},\textbf{132770},,
plus63mod4096\_163,56329,48265,--,--,--,--,--,--,\textbf{155209},\textbf{135715},,
urf5\_158,71932,64750,--,--,--,--,--,--,\textbf{177757},\textbf{158774},,
urf6\_160,75180,67637,--,--,--,--,--,--,\textbf{209231},\textbf{174548},,
urf1\_149,80878,72933,--,--,--,--,--,--,\textbf{200949},\textbf{179071},,
plus63mod8192\_164,81865,70222,--,--,--,--,--,--,\textbf{226354},\textbf{198162},,
hwb9\_119,90955,77968,--,--,--,--,--,--,\textbf{233049},\textbf{205682},,
urf3\_155,185276,167215,--,--,--,--,--,--,\textbf{463538},\textbf{409727},,
ground\_state \_estimation\_10,154209,151095,--,--,--,--,--,--,\textbf{5039},\textbf{3897},,
urf4\_187,224028,185975,--,--,--,--,--,--,\textbf{568938},\textbf{481368},,
\end{filecontents*}
\csvreader[
    respect underscore, 
    late after line=\\,
    late after last line=,
    ]{benchmarkQX5CX.csv}
    {1=\Name, 2=\Gates, 3=\Depth, 4=\gIBM, 5=\dIBM, 6=\gProQ, 7=\dProQ, 8=\gRig, 9=\dRig, 10=\gCQC, 11=\dCQC, 12=\sTketBest, 13=\dTketBest}
    {\Name  &  \Gates & \Depth & \gIBM & \dIBM  & \gProQ & \dProQ & \gRig & \dRig & \gCQC & \dCQC &  \sTketBest & \dTketBest }
      \end{longtable*}
  }

\clearpage
\subsection{CX only comparison on IBM Tokyo}
\label{sec:cx-only-comparison}

\begin{figure}[h]
  \centering
\includegraphics[width=.5\linewidth]{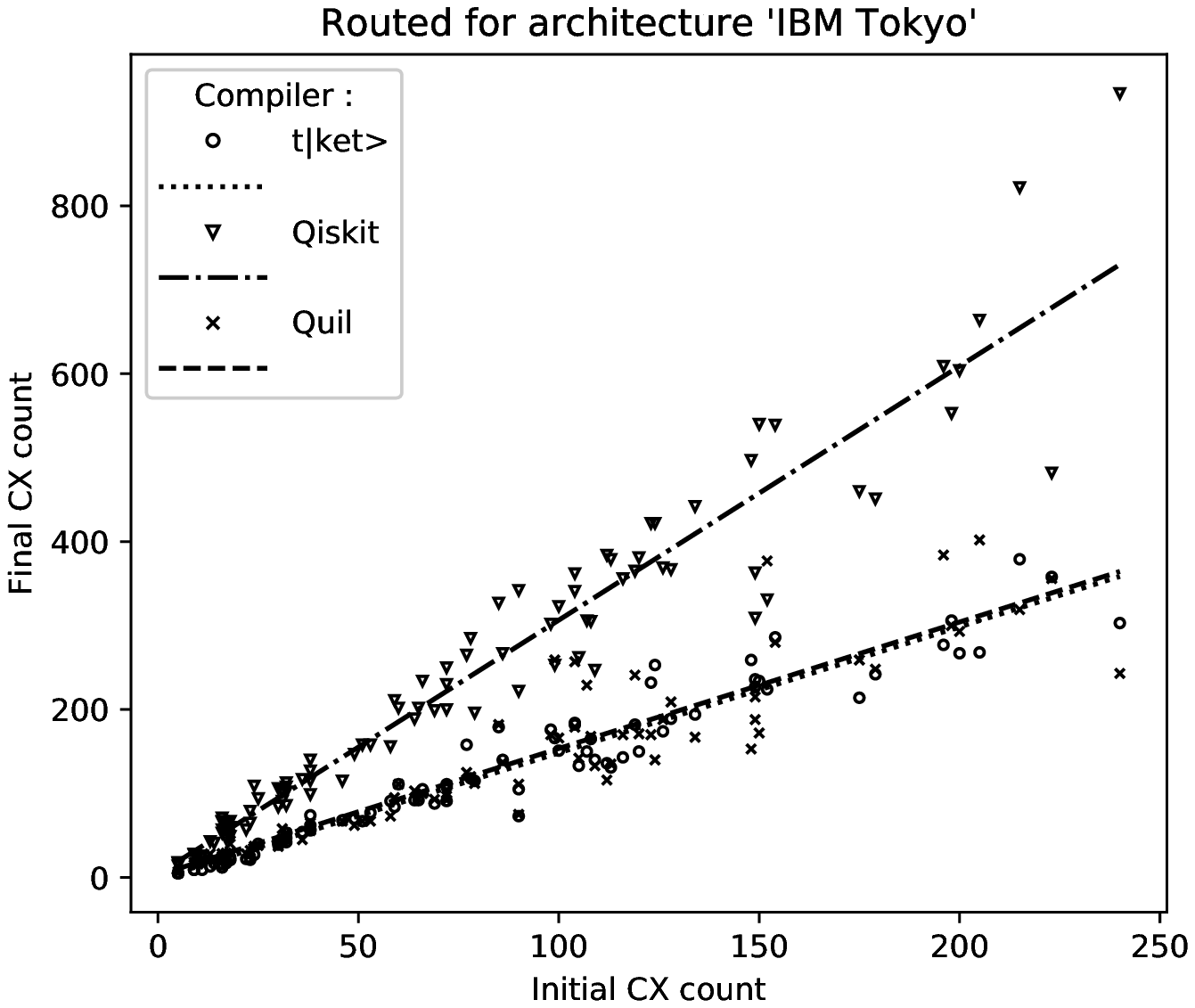}%
\includegraphics[width=.5\linewidth]{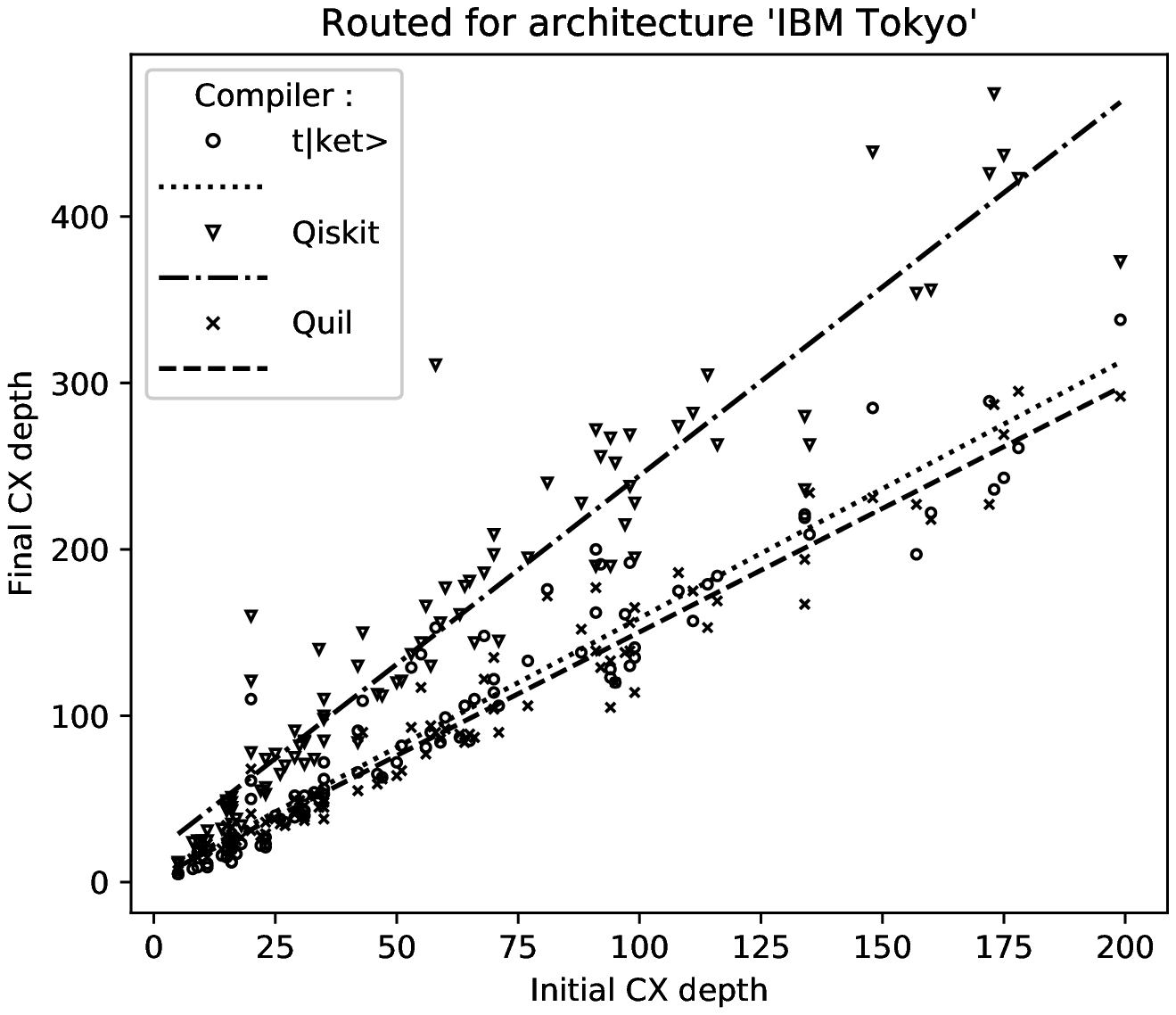}%

\caption{Routing comparison on IBM Tokyo, CX count and CX depth when counting only CX gates.  The  charts are a zoomed in version of the initial segment of lower charts of Fig.~\ref{fig:compare}.}
\end{figure}

  {\setlength{\tabcolsep}{0.09cm}
      \begin{longtable*}{p{3.5cm}rr||rr|rr||rr|rr} 
        \caption{CX gates only comparison on IBM Tokyo}
        \label{tab:results-cx-tokyo}\\ 
    \multicolumn{3}{c||}{}  & \multicolumn{2}{c|}{\makecell{Qiskit \\ 0.7.0}}  & \multicolumn{2}{c||}{\makecell{Quilc 1.1.1 \\ Pyquil 2.1.1}}& \multicolumn{2}{c}{CQC's \tket}&\multicolumn{2}{|c}{\makecell{\tket \\  comparison }} \\
    Name  & $g_{\rm in}$ & $d_{\rm in}$ & $g_{\rm out}$ & $d_{\rm out}$ & $g_{\rm out}$ & $d_{\rm out}$ &  $g_{\rm out}$ &$d_{\rm out}$ & $r_{\rm gate}$ & $r_{\rm depth}$ \\ \hline  \endfirsthead
    
    \multicolumn{3}{c||}{}  & \multicolumn{2}{c|}{\makecell{Qiskit \\ 0.7.0}} & \multicolumn{2}{c||}{\makecell{Quilc 1.1.1 \\ Pyquil 2.1.1}}& \multicolumn{2}{c}{CQC's \tket}&\multicolumn{2}{|c}{\makecell{\tket  \\  comparison}} \\
    Name  & $g_{\rm in}$ & $d_{\rm in}$ & $g_{\rm out}$ & $d_{\rm out}$ & $g_{\rm out}$ & $d_{\rm out}$ &  $g_{\rm out}$ &$d_{\rm out}$ & $r_{\rm gate}$ & $r_{\rm depth}$ \\ \hline  \endhead
    
    \hline
\endfoot    

    \hline \\
\multicolumn{11}{l}{ $g$: the number of quantum gates (elementary operations),}\\
\multicolumn{11}{l}{$d$: depth of the quantum circuits and  -- are time-outs.}
\endlastfoot
        \begin{filecontents*}{benchmarkTokyoCX.csv}
name,Size in,Depth in,CX_Size_qiskit,CX_Depth_qiskit,CX_Size_rigetti,CX_Depth_rigetti,CX_Size_CQC,CX_Depth_CQC,Size_ratioTketNextBest,Depth_ratioTketNextBest
xor5\_254,5,5,14,11,7,7,\textbf{5},\textbf{5},0.71,0.71
graycode6\_47,5,5,17,11,\textbf{5},\textbf{5},\textbf{5},\textbf{5},1.00,1.00
ex1\_226,5,5,18,12,7,7,\textbf{5},\textbf{5},0.71,0.71
4gt11\_84,9,8,27,24,15,14,\textbf{9},\textbf{8},0.60,0.57
4mod5-v0\_20,10,9,25,19,\textbf{16},\textbf{15},19,18,1.19,1.20
ex-1\_166,9,9,28,25,18,15,\textbf{9},\textbf{9},0.50,0.60
4mod5-v1\_22,11,10,26,25,\textbf{20},\textbf{18},23,22,1.15,1.22
mod5d1\_63,13,11,43,31,28,23,\textbf{13},\textbf{11},0.46,0.48
ham3\_102,11,11,25,25,18,15,\textbf{9},\textbf{9},0.50,0.60
4gt11\_83,14,14,41,32,20,20,\textbf{17},\textbf{16},0.85,0.80
4gt11\_82,18,18,49,34,30,27,\textbf{24},\textbf{23},0.80,0.85
rd32-v0\_66,16,16,57,43,21,20,\textbf{12},\textbf{12},0.57,0.60
alu-v0\_27,17,15,63,48,53,34,\textbf{20},\textbf{19},0.38,0.56
4mod5-v1\_24,16,15,66,49,28,27,\textbf{16},\textbf{15},0.57,0.56
4mod5-v0\_19,16,16,67,51,\textbf{25},\textbf{25},\textbf{25},\textbf{25},1.00,1.00
mod5mils\_65,16,16,71,48,28,28,\textbf{25},\textbf{25},0.89,0.89
rd32-v1\_68,16,16,53,45,24,17,\textbf{12},\textbf{12},0.50,0.71
alu-v1\_28,18,16,60,35,40,26,\textbf{21},\textbf{20},0.53,0.77
alu-v2\_33,17,15,59,44,29,28,\textbf{20},\textbf{19},0.69,0.68
alu-v4\_37,18,16,67,45,52,33,\textbf{21},\textbf{20},0.40,0.61
alu-v3\_35,18,16,57,43,52,33,\textbf{21},\textbf{20},0.40,0.61
3\_17\_13,17,17,46,38,26,25,\textbf{17},\textbf{17},0.65,0.68
alu-v1\_29,17,15,57,43,53,34,\textbf{20},\textbf{19},0.38,0.56
miller\_11,23,23,65,53,32,29,\textbf{23},\textbf{23},0.72,0.79
alu-v3\_34,24,23,109,74,37,36,\textbf{27},\textbf{27},0.73,0.75
decod24-v2\_43,22,22,57,55,28,28,\textbf{22},\textbf{22},0.79,0.79
decod24-v0\_38,23,23,79,57,27,25,\textbf{21},\textbf{21},0.78,0.84
mod5d2\_64,25,25,94,77,\textbf{38},\textbf{38},40,40,1.05,1.05
4gt13\_92,30,26,84,65,\textbf{39},\textbf{35},42,38,1.08,1.09
4gt13-v1\_93,30,27,106,70,\textbf{37},\textbf{34},39,36,1.05,1.06
4mod5-v0\_18,31,31,91,71,58,46,\textbf{43},\textbf{40},0.74,0.87
decod24-bdd\_294,32,31,102,85,\textbf{44},\textbf{37},53,52,1.20,1.41
one-two-three-v2\_100,32,29,86,75,\textbf{46},\textbf{41},53,52,1.15,1.27
one-two-three-v3\_101,32,29,108,91,53,50,\textbf{42},\textbf{39},0.79,0.78
4mod5-v1\_23,32,30,113,82,50,49,\textbf{47},\textbf{42},0.94,0.86
4mod5-bdd\_287,31,31,103,84,46,\textbf{39},\textbf{43},43,0.93,1.10
rd32\_270,36,35,117,98,\textbf{45},\textbf{38},54,53,1.20,1.39
4gt5\_75,38,33,99,74,\textbf{56},\textbf{53},\textbf{56},54,1.00,1.02
alu-bdd\_288,38,35,140,110,\textbf{65},\textbf{45},74,72,1.14,1.60
alu-v0\_26,38,35,115,85,\textbf{56},\textbf{48},57,56,1.02,1.17
decod24-v1\_41,38,35,127,100,\textbf{56},\textbf{50},62,62,1.11,1.24
rd53\_138,60,42,202,130,\textbf{111},\textbf{87},\textbf{111},91,1.00,1.05
4gt5\_76,46,42,115,84,\textbf{67},\textbf{55},68,66,1.01,1.20
4gt13\_91,49,46,147,113,\textbf{62},\textbf{59},70,65,1.13,1.10
cnt3-5\_179,85,43,327,150,182,\textbf{90},\textbf{179},109,0.98,1.21
qft\_10,90,34,342,140,75,\textbf{45},\textbf{73},50,0.97,1.11
4gt13\_90,53,50,158,120,\textbf{67},\textbf{64},77,72,1.15,1.12
alu-v4\_36,51,47,158,112,\textbf{67},\textbf{62},\textbf{67},63,1.00,1.02
mini\_alu\_305,77,53,265,137,\textbf{125},\textbf{93},158,129,1.26,1.39
ising\_model\_10,90,20,222,78,111,\textbf{41},\textbf{105},50,0.95,1.22
ising\_model\_16,150,20,540,160,\textbf{172},\textbf{31},234,110,1.36,3.55
ising\_model\_13,120,20,381,121,171,68,\textbf{150},\textbf{61},0.88,0.90
4gt5\_77,58,51,156,121,\textbf{73},\textbf{67},91,82,1.25,1.22
sys6-v0\_111,98,55,302,144,\textbf{170},\textbf{117},176,137,1.04,1.17
one-two-three-v1\_99,59,56,211,166,95,\textbf{77},\textbf{84},81,0.88,1.05
one-two-three-v0\_98,65,59,202,156,94,86,\textbf{92},\textbf{84},0.98,0.98
decod24-v3\_45,64,57,189,130,103,94,\textbf{92},\textbf{90},0.89,0.96
4gt10-v1\_81,66,60,234,177,\textbf{100},\textbf{92},105,99,1.05,1.08
aj-e11\_165,69,63,199,161,93,89,\textbf{88},\textbf{87},0.95,0.98
4mod7-v0\_94,72,66,200,144,\textbf{97},\textbf{87},111,110,1.14,1.26
alu-v2\_32,72,64,250,178,\textbf{92},\textbf{84},109,106,1.18,1.26
rd73\_140,104,68,341,186,\textbf{179},\textbf{122},182,148,1.02,1.21
4mod7-v1\_96,72,65,230,181,100,89,\textbf{91},\textbf{85},0.91,0.96
4gt4-v0\_80,79,71,196,145,\textbf{112},\textbf{90},115,106,1.03,1.18
mod10\_176,78,70,285,197,120,\textbf{104},\textbf{118},114,0.98,1.10
0410184\_169,104,70,362,209,257,135,\textbf{184},\textbf{122},0.72,0.90
qft\_16,240,58,934,311,\textbf{243},\textbf{90},303,153,1.25,1.70
4gt12-v0\_88,86,77,267,195,\textbf{137},\textbf{106},140,133,1.02,1.25
rd84\_142,154,81,539,240,\textbf{280},\textbf{172},286,176,1.02,1.02
rd53\_311,124,92,422,256,--,--,\textbf{253},\textbf{191},0.60,0.75
4\_49\_16,99,91,253,190,\textbf{140},\textbf{129},166,162,1.19,1.26
sym9\_146,148,91,497,272,\textbf{259},\textbf{177},\textbf{259},200,1.00,1.13
4gt12-v1\_89,100,88,323,228,153,139,\textbf{151},\textbf{138},0.99,0.99
4gt12-v0\_87,112,94,384,267,166,152,\textbf{136},\textbf{123},0.82,0.81
4gt4-v0\_79,105,94,262,190,\textbf{116},\textbf{105},133,128,1.15,1.22
hwb4\_49,107,99,306,228,\textbf{142},\textbf{133},150,141,1.06,1.06
sym6\_316,123,98,422,269,\textbf{229},\textbf{165},232,192,1.01,1.16
4gt12-v0\_86,116,98,356,238,170,156,\textbf{143},\textbf{130},0.84,0.83
4gt4-v0\_72,113,95,379,252,170,139,\textbf{131},\textbf{120},0.77,0.86
4gt4-v0\_78,109,99,247,195,\textbf{135},\textbf{120},140,135,1.04,1.12
mod10\_171,108,97,305,215,\textbf{133},\textbf{114},165,161,1.24,1.41
4gt4-v1\_74,119,108,365,274,\textbf{168},\textbf{138},182,175,1.08,1.27
rd53\_135,134,114,442,305,241,186,\textbf{194},\textbf{179},0.80,0.96
mini-alu\_167,126,111,369,282,\textbf{167},\textbf{153},174,157,1.04,1.03
one-two-three-v0\_97,128,116,367,263,\textbf{188},\textbf{175},189,184,1.01,1.05
ham7\_104,149,134,309,236,\textbf{209},\textbf{169},236,221,1.13,1.31
decod24-enable\_126,149,134,363,280,\textbf{215},\textbf{194},227,219,1.06,1.13
mod8-10\_178,152,135,331,263,\textbf{188},\textbf{167},224,209,1.19,1.25
cnt3-5\_180,215,148,822,439,\textbf{377},\textbf{234},379,285,1.01,1.22
ex3\_229,175,157,460,354,319,231,\textbf{214},\textbf{197},0.67,0.85
4gt4-v0\_73,179,160,451,356,259,227,\textbf{242},\textbf{222},0.93,0.98
mod8-10\_177,196,178,609,423,\textbf{248},\textbf{218},277,261,1.12,1.20
C17\_204,205,173,664,474,384,295,\textbf{268},\textbf{236},0.70,0.80
alu-v2\_31,198,172,553,426,402,\textbf{287},\textbf{306},289,0.76,1.01
rd53\_131,200,175,604,437,300,\textbf{227},\textbf{267},243,0.89,1.07
alu-v2\_30,223,199,482,373,\textbf{293},\textbf{269},358,338,1.22,1.26
mod5adder\_127,239,208,650,476,356,292,\textbf{311},\textbf{288},0.87,0.99
rd53\_133,256,221,703,499,359,\textbf{308},\textbf{348},328,0.97,1.06
cm82a\_208,283,234,914,607,\textbf{393},\textbf{317},451,391,1.15,1.23
majority\_239,267,232,720,528,402,328,\textbf{348},\textbf{311},0.87,0.95
ex2\_227,275,241,886,648,443,\textbf{371},\textbf{416},391,0.94,1.05
sf\_276,336,301,989,731,\textbf{372},\textbf{340},489,453,1.31,1.33
sf\_274,336,300,863,643,457,391,\textbf{357},\textbf{328},0.78,0.84
con1\_216,415,346,1389,916,\textbf{659},\textbf{486},736,626,1.12,1.29
wim\_266,427,352,1145,793,\textbf{763},\textbf{501},772,688,1.01,1.37
rd53\_130,448,383,1242,899,695,\textbf{547},\textbf{646},608,0.93,1.11
f2\_232,525,449,1443,1069,\textbf{722},\textbf{621},854,772,1.18,1.24
cm152a\_212,532,461,1491,1051,805,\textbf{633},\textbf{793},717,0.99,1.13
rd53\_251,564,492,1487,1058,960,\textbf{659},\textbf{805},715,0.84,1.08
hwb5\_53,598,535,1568,1179,941,765,\textbf{808},\textbf{757},0.86,0.99
cm42a\_207,771,634,2333,1593,1411,996,\textbf{1102},\textbf{968},0.78,0.97
pm1\_249,771,634,2303,1574,1411,996,\textbf{1102},\textbf{968},0.78,0.97
dc1\_220,833,705,2450,1715,1532,\textbf{1049},\textbf{1436},1261,0.94,1.20
squar5\_261,869,720,2902,1948,\textbf{1463},\textbf{1063},1631,1372,1.11,1.29
z4\_268,1343,1112,4185,2828,2282,\textbf{1632},\textbf{2235},1914,0.98,1.17
sqrt8\_260,1314,1121,4079,2846,2406,\textbf{1665},\textbf{2235},1964,0.93,1.18
radd\_250,1405,1210,4346,3050,--,--,\textbf{2697},\textbf{2361},0.62,0.77
adr4\_197,1498,1249,4584,3110,\textbf{2333},\textbf{1720},2737,2368,1.17,1.38
sym6\_145,1701,1499,4872,3517,2554,\textbf{2042},\textbf{2301},2145,0.90,1.05
misex1\_241,2100,1797,--,--,\textbf{3249},\textbf{2435},4291,3760,1.32,1.54
rd73\_252,2319,1963,--,--,3854,\textbf{2798},\textbf{3748},3276,0.97,1.17
cycle10\_2\_110,2648,2276,--,--,4511,\textbf{3463},\textbf{4342},3802,0.96,1.10
hwb6\_56,2952,2559,--,--,4405,\textbf{3575},\textbf{4205},3904,0.95,1.09
square\_root\_7,3089,2520,--,--,\textbf{4967},\textbf{3427},5790,4833,1.17,1.41
ham15\_107,3858,3273,--,--,6649,\textbf{4828},\textbf{6485},5605,0.98,1.16
dc2\_222,4131,3518,--,--,--,--,\textbf{7694},\textbf{6708},,
sqn\_258,4459,3719,--,--,--,--,\textbf{6863},\textbf{5984},,
inc\_237,4636,3928,--,--,--,--,\textbf{8274},\textbf{7176},,
cm85a\_209,4986,4256,--,--,--,--,\textbf{8166},\textbf{7093},,
rd84\_253,5960,4917,--,--,--,--,\textbf{9506},\textbf{8154},,
co14\_215,7840,5759,--,--,--,--,\textbf{14415},\textbf{11195},,
root\_255,7493,5965,--,--,--,--,\textbf{11971},\textbf{9940},,
mlp4\_245,8232,6930,--,--,--,--,\textbf{15156},\textbf{13061},,
urf2\_277,10066,8312,--,--,--,--,\textbf{17367},\textbf{15384},,
sym9\_148,9408,8062,--,--,--,--,\textbf{13893},\textbf{12321},,
life\_238,9800,8356,--,--,--,--,\textbf{15944},\textbf{13926},,
hwb7\_59,10681,9112,--,--,--,--,\textbf{17655},\textbf{15828},,
max46\_240,11844,9657,--,--,--,--,\textbf{19519},\textbf{16979},,
clip\_206,14772,12028,--,--,--,--,\textbf{25499},\textbf{21345},,
9symml\_195,15232,12849,--,--,--,--,\textbf{24532},\textbf{21295},,
sym9\_193,15232,12849,--,--,--,--,\textbf{24532},\textbf{21295},,
sao2\_257,16864,13209,--,--,--,--,\textbf{28799},\textbf{23847},,
dist\_223,16624,13274,--,--,--,--,\textbf{27953},\textbf{23342},,
urf5\_280,23764,19888,--,--,--,--,\textbf{41610},\textbf{36329},,
urf1\_278,26692,22307,--,--,--,--,\textbf{46023},\textbf{40020},,
sym10\_262,28084,23736,--,--,--,--,\textbf{44383},\textbf{38132},,
hwb8\_113,30372,26041,--,--,--,--,\textbf{50008},\textbf{44245},,
urf2\_152,35210,32247,--,--,--,--,\textbf{58307},\textbf{53905},,
urf3\_279,60380,50568,--,--,--,--,\textbf{105870},\textbf{91838},,
plus63mod4096\_163,56329,48265,--,--,--,--,\textbf{95107},\textbf{82375},,
urf5\_158,71932,64750,--,--,--,--,\textbf{122288},\textbf{110318},,
urf6\_160,75180,67637,--,--,--,--,\textbf{141416},\textbf{120999},,
urf1\_149,80878,72933,--,--,--,--,\textbf{137791},\textbf{123873},,
plus63mod8192\_164,81865,70222,--,--,--,--,\textbf{145872},\textbf{125957},,
hwb9\_119,90955,77968,--,--,--,--,\textbf{149106},\textbf{131678},,
urf3\_155,185276,167215,--,--,--,--,\textbf{317732},\textbf{285000},,
ground\_state \_estimation\_10,154209,151095,--,--,--,--,\textbf{1015},\textbf{776},,
urf4\_187,224028,185975,--,--,--,--,\textbf{385355},\textbf{330368},,
\end{filecontents*}
\csvreader[
    respect underscore, 
    late after line=\\,
    late after last line=,
    ]{benchmarkTokyoCX.csv}
    {1=\Name, 2=\Gates, 3=\Depth, 4=\gIBM, 5=\dIBM, 6=\gRig, 7=\dRig, 8=\gCQC, 9=\dCQC, 10=\sTketBest, 11=\dTketBest}
    {\Name  &  \Gates & \Depth & \gIBM & \dIBM & \gRig & \dRig & \gCQC & \dCQC &  \sTketBest & \dTketBest }
      \end{longtable*}
  }

    \end{document}